%% file: article.tex
\documentclass[twocolumn, twocolappendix]{aastex631}

\usepackage{CJK}
\usepackage[dvipsnames]{xcolor}
\usepackage{hyperref}
\usepackage{subfigure}
\usepackage{underscore}
\usepackage{multirow}
\usepackage{makecell}
\usepackage{amsmath}
\usepackage[T1]{fontenc}

\newcommand{\diskname}{MWC 480}

\newcommand{\uat}[2]{\href{http://astrothesaurus.org/uat/#1}{#2 (#1)}}

\begin{document}
\begin{CJK*}{UTF8}{gkai}

\title{Probing Dust in the MWC 480 Disk from Millimeter to Centimeter Wavelengths}

\correspondingauthor{Yangfan Shi, Feng Long, Enrique Mac\'{i}as, Gregory Herczeg}

\author[0000-0001-9277-6495]{Yangfan Shi (施杨帆)}
\affiliation{Kavli Institute for Astronomy and Astrophysics, Peking University, Beijing 100871, China}
\affiliation{Department of Astronomy, Peking University, Beijing 100871, China}
\affiliation{European Southern Observatory, Karl-Schwarzschild-Str. 2, D-85748 Garching bei M\"{u}nchen, Germany}
\email{yfshi.astro@gmail.com}

\author[0000-0002-7607-719X]{Feng Long (龙凤)}
\altaffiliation{NASA Hubble Fellowship Program Sagan Fellow}
\affiliation{Kavli Institute for Astronomy and Astrophysics, Peking University, Beijing 100871, China}
\affiliation{Department of Astronomy, Peking University, Beijing 100871, China}
\affiliation{Lunar and Planetary Laboratory, University of Arizona, Tucson, AZ 85721, USA}
\email{long.feng@pku.edu.cn}

\author[0000-0003-1283-6262]{Enrique, Mac\'{i}as}
\affiliation{European Southern Observatory, Karl-Schwarzschild-Str. 2, D-85748 Garching bei M\"{u}nchen, Germany}
\email{enrique.macias@eso.org}

\author[0000-0002-7154-6065]{Gregory J. Herczeg (沈雷歌)}
\affiliation{Kavli Institute for Astronomy and Astrophysics, Peking University, Beijing 100871, China}
\affiliation{Department of Astronomy, Peking University, Beijing 100871, China}
\email{gherczeg1@gmail.com}

\author[0000-0001-8764-1780]{Paola Pinilla}
\affiliation{Mullard Space Science Laboratory, University College London, Holmbury St Mary, Dorking, Surrey RH5 6NT, UK}

\author[0000-0003-2253-2270]{Sean M. Andrews}
\affiliation{Center for Astrophysics, Harvard \& Smithsonian, 60 Garden Street, Cambridge, MA 02138, USA}

\author[0000-0003-1526-7587]{David J. Wilner}
\affiliation{Center for Astrophysics, Harvard \& Smithsonian, 60 Garden Street, Cambridge, MA 02138, USA}

\author[0000-0003-2948-5614]{Haochang Jiang (蒋昊昌)}
\affiliation{Max-Planck-Institut für Astronomie, Königstuhl 17, 69117 Heidelberg, Germany}

\author[0000-0001-9290-7846]{Ruobing Dong (董若冰)}
\affiliation{Kavli Institute for Astronomy and Astrophysics, Peking University, Beijing 100871, China}
\affiliation{Department of Astronomy, Peking University, Beijing 100871, China}

\author[0000-0003-1534-5186]{Richard Teague}
\affiliation{Department of Earth, Atmospheric, and Planetary Sciences, Massachusetts Institute of Technology, Cambridge, MA 02139, USA}

\author[0000-0001-7962-1683]{Ilaria Pascucci}
\affiliation{Lunar and Planetary Laboratory, The University of Arizona, Tucson, AZ 85721, USA}

\author[0000-0002-6958-4986]{Claudia Toci}
\affiliation{Escuela Técnica Superior de Ingeniería, Universidad de Sevilla, Camino de los Descubrimientos s/n, 41092 Sevilla, Spain}

\author[0000-0003-3283-6884]{Yuri Aikawa}
\affiliation{Department of Applied Physics, Faculty of Engineering, Kanagawa University, Kanagawa 221-0802, Japan}

\author[0000-0001-6307-4195]{Daniel Harsono}
\affiliation{Institute of Astronomy, Department of Physics, National Tsing Hua University, Hsinchu 30013, Taiwan}

\author[0000-0002-7616-666X]{Yao Liu (刘尧)}
\affiliation{School of Physical Science and Technology, Southwest Jiaotong University, Chengdu 610031, China}

\begin{abstract}

We present deep, high-resolution ($\sim$100 mas) Karl G. Jansky Very Large Array (VLA) Ka-band (9.1 mm) observations of the disk around MWC 480, and infer dust properties through a combined analysis with archival Atacama Large Millimeter/submillimeter Array (ALMA) data at 0.87, 1.17, 1.33, and 3.0 mm.
The prominent dust ring at 95 au (B95) is detected at 9.1 mm for the first time, while the faint outer ring at 160 au is not revealed.
Through non-parametric visibility modeling, we identified two new annular features: a plateau within 20-50 au across all wavelengths, and a shoulder exterior to the B95 ring at 0.87, 1.17 and 1.33 mm, consistent with signatures of planet-disk interaction.
We find that the width of the B95 ring remains constant across wavelengths, suggesting that fragmentation dominates over radial diffusion or that unresolved substructure is present within the ring.
Resolved spectral modeling yields two families of dust solutions that reproduce the observations equally well: compact grains or highly porous (90\%) grains, with carbonaceous components dominated by refractory organics or amorphous carbon, respectively.
The inferred maximum grain sizes peak at the locations of the two rings and reach centimeter within the B95 ring.
The total dust masses are $860^{+95}_{-78}\rm~M_\oplus$/$1500^{+440}_{-330}\rm~M_\oplus$ (large/small-grain solution in inner disk) and $230^{+14}_{-13}\rm~M_\oplus$ for the two dust mixtures. 
The B95 ring alone contains $100^{+5}_{-5}\rm~M_\oplus$ and $43^{+2}_{-2}\rm~M_\oplus$, respectively, sufficient to assemble the cores of giant planets.
Finally, we highlight the power of broadband, multi-wavelength observations in placing better constraints on dust composition and porosity in protoplanetary disks.

\end{abstract}

\keywords{\uat{1300}{Protoplanetary disks}; \uat{1241}{Planet Formation};
\uat{236}{Circumstellar dust};
\uat{2204}{Planetary-disk interactions};
\uat{1346}{Radio interferometry}}

\section{Introduction} \label{sec:intro}

Dust particles are the building blocks of planets.
On the way to forming planets, these particles must grow from sub-micrometer sizes inherited from ISM \citep{mathis1977} to kilometer-sized objects.
As dust grows, several barriers arise due to bouncing, erosion, and fragmentation, as the relative collision speeds increase with particle size. The radial drift barrier is also expected once the inward drift is faster than growth \citep[reviews by][and references therein]{testi14_pp, birnstiel24_araa}.

The advent of the Atacama Large Millimeter/submillimeter Array (ALMA) has revolutionized our view of protoplanetary disks, revealing that many disks are structured on scales of a few au \citep[reviews by][]{andrews20_araa, bae23}.
The most common substructures are in the form of rings and gaps \citep[e.g.,][]{hltau15, andrews18, long18, huang18, cieza21, kurtovic21, shi24}, often interpreted as evidence of efficient dust trapping \citep[e.g.,][]{dullemond18, rosotti20, doi_kataoka23, carvalho24, sierra25}.
Such dust traps can concentrate solids, enabling them to overcome growth barriers and potentially trigger planetesimal formation via streaming instability \citep{youdin_goodman05, stammler19, li_youdin21}.
Characterization of dust properties in the rings such as mass and size at high resolution is therefore essential to understand dust evolution and further planet formation.

Multi-wavelength continuum observations at (sub)millimeter to centimeter wavelengths provide an efficient tool to study dust in disks.
The method relies on the frequency dependence of dust opacity ($\kappa_\nu \propto \nu^\beta$), where the opacity index $\beta$ encodes information about the grain properties \citep{beckwith1990}.
Larger dust particles tend to yield lower $\beta$ values, and thus a flatter spectrum \citep{miyake93, dlessio01}.
However, high optical depths can also flatten the spectrum, creating a degeneracy between large particle sizes and high optical depths.
This degeneracy can be mitigated by combining observations that span a wide range of optical depths, ideally covering both optically thick and thin regimes.

In recent years, dust analysis using high resolution multi-wavelength continuum observations has been carried out with ALMA \citep[e.g.,][]{macias19, macias21, sierra21, ueda22, carvalho24, sierra25}, sometimes combined with the Karl G. Jansky Very Large Array (VLA) at longer wavelength \citep[e.g.,][]{carrasco-gonzalez19, guidi22, zhang23_hltau, sierra24, ueda25_hltau, zagaria25}.
The nearly order-of-magnitude span in wavelengths allows for revealing the intrinsic dust properties from high optical depth.
However, the constraints on dust properties can vary significantly across studies, sometimes even for the same disk.
It is because different assumptions are often adopted in dust mixture models, such as composition and porosity, which strongly affect the opacities \citep[e.g.,][]{birnstiel18} and consequently the inferred dust size and mass.
While different dust mixtures can reproduce the fluxes at a few millimeters (most commonly from ALMA Band 7/6 to Band 3) reasonably well, they may diverge significantly at longer wavelength \citep[e.g.,][]{sierra25}. 
A broad wavelength coverage therefore provides an opportunity to test different dust mixtures \citep[e.g.,][]{guidi22, zhang23_hltau, zagaria25}, or even to treat the dust mixture as a free parameter to fit \citep[][for the HL Tau disk]{ueda25_hltau}.

In this study, we carry out a multi-wavelength analysis of continuum emission to probe the dust properties in the MWC 480 protoplanetary disk.
MWC 480 is a $\sim7$ Myr old Herbig Ae star with a dynamical mass of $2.1\rm~M_\odot$ \citep{simon19}.
Its disk exhibits a bright inner smooth core, and two rings at $\sim98$ and $\sim165$ au \citep{long18, liu19, law21, sierra21}.
Previously, \citet{sierra21} performed multi-wavelength continuum analysis on the MWC 480 disk using observations at ALMA Band 6 and 3 (257/226/100 GHz), with a fixed dust temperature profile from \citet{zhang21_maps}.
Due to the limited sensitivity and non-detection of the $\sim160$ au ring at Band 3, the fitting is performed at a resolution $\sim 0.22''$ and only covers the radii out to the ring at $\sim100$ au.
They have retrieved radially decreasing maximum grain sizes of $\sim4$ mm in the inner disk and $\sim 2$ mm in the outer disk, the overall profile is rather flat and does not show a local maximum at the ring position.

We present deep VLA observations of the disk around MWC 480 (HD 31648) in Ka band ($\sim9\rm~mm$) at a resolution $\sim0.1''$, and further collect archival data at ALMA Bands 7/6/3 to revisit the properties of dust in the MWC 480 disk.
We note here that GAIA DR3 \citep{gaiadr3} reports a parallax $6.40\pm0.05$ mas, larger than $6.18\pm0.08$ mas by GAIA DR2 \citep{gaiadr2}.
We adopt the updated parallax distance to MWC 480 of 156.2 pc, which shifts two rings to $\sim$ 95 and 160 au, and update its derived physical properties involving distance when needed, such as stellar luminosity. 

We organize this paper as follows:
In Section \ref{sec:obs}, we present VLA observations of MWC 480 disk, and describe the collection and concatenation of ALMA observations at Band 7/6/3.
We extract observables from continuum images and radial profiles from modeling visibilities, and compare them across wavelength in Section \ref{sec:obs_results}.
In Section \ref{sec:model_dust}, we investigate the dust properties as a function of radius by fitting the multi-wavelength radial profiles.
We discuss the results and their implications in Section \ref{sec:discussion} and conclude in Section \ref{sec:summary}.

\section{Observations} \label{sec:obs}
We introduce our new VLA observations at Ka and X band. For MWC 480 at shorter mm wavelengths, we make use of ALMA archival data at Band 7/6/3.
The programs at all wavelengths used in this paper are summarized in Table \ref{tab:data_collection}.

\begin{deluxetable*}{lcccccc}
    \label{tab:data_collection}
    \tablecaption{Summary of multi-wavelength programs used in this work.}
    \tablehead{
    \colhead{Band} & \colhead{Project code} & \colhead{PI} & \colhead{Frequency range} & \colhead{On-source time} & \colhead{Baselines} & \colhead{MRS\tablenotemark{a}} \\
    \colhead{} & \colhead{} & \colhead{} & \colhead{(GHz)} & \colhead{(mins)} & \colhead{(m)} & \colhead{(arcsec)}
    }
    \startdata
    ALMA/Band 7 & 2017.1.00470.S & Looney, L. & 335.5 - 351.5 & 24.19 m & 15.1 - 1397.8 & 2.54 \\[0.08cm] \hline
    \multirow{2}{*}{ALMA/Band 6} & \multirow{2}{*}{2018.1.01055.L} & \multirow{2}{*}{Oberg, K.} & 248.1 - 266.0 & 218.27 m & 15.1 - 5893.6 & 2.39 \\[0.08cm] \cline{4-7}
      &   &   & 217.2 - 234.9 & 263.35 m & 15.1 - 3637.7 & 2.65 \\[0.08cm] \hline
    \multirow{3}{*}{ALMA/Band 3} & \multirow{2}{*}{2018.1.01055.L} & \multirow{2}{*}{Oberg, K.} & 97.9 - 113.5 & 215.73 m & 15.1 - 3637.7 & 7.83 \\ \cline{4-7}
     &  &  & 86.0 - 101.4 & 150.88 m & 15.1 - 3637.7 & 8.22 \\ \cline{2-7}
     & 2016.1.01042.S & Chandler, C. & 94.5 - 110.3 & 3.2 m & 41.4 - 12145.2 & 1.65 \\[0.08cm] \hline
    \multirow{3}{*}{VLA/Ka} & 21B$-$141 & Long, F. & 29 - 37 & 5.3 h & 243.1 - 11128.0 & 2.86 \\[0.08cm] \cline{2-7}
     &  20B$-$342 & Long, F. & 29 - 37 & 16.3 h & 498.3 - 36623.1 & 0.98 \\[0.08cm] \hline
    \multirow{2}{*}{VLA/X} &  21B$-$141 & Long, F. & 8 - 12 & 24.2 m & 243.1 - 11128.0 & 9.44 \\ \cline{2-7}
     &  20B$-$342 & Long, F. & 8 - 12 & 81.6 m & 256.4 - 36623.1 & 3.94 \\
    \enddata
    \tablenotetext{a}{Maximum Recoverable Scale.}
    \tablecomments{Information retrieved using \texttt{analysisUtils} \citep{analysisUtils}.}
\end{deluxetable*}

\subsection{VLA Observations \& Calibration} 
\label{subsec:obs_vla}
\paragraph{VLA Ka Band~}
~We obtained deep VLA observations of \diskname~at Ka ($\sim$ 9 mm)  band through projects 20B-342 and 21B-141 (PI: Feng Long). 
The frequency ranged between 29 and 37 GHz.
The A array observations were conducted in 2021 and 2022 through project 20B-342, consisting of 15 execution blocks between 2021-Jan-02 and 2021-Feb-27 and 10 execution blocks between 2022-Apr-25 and 2022-Jun-14. The B array observations were conducted in 2021 with 7 execution blocks between 2021-Oct-03 and 2021-Oct-20 via project 21B-141. 
The total on-source time adds up to 16.3 and 5.3 hours for A and B array observations, respectively.
The quasar 3C147 was used as the flux and bandpass calibrator, while J0443+3441 acted as the complex gain calibrator.

\paragraph{VLA X Band~}
~The two VLA projects were also accompanied with X ($\sim$3 cm) band observations with frequency range between 8 and 12 GHz. 
The total integration time on science target is 69.6 mins in A array, 11.9 mins in BnA$\rightarrow$A, and 24.2 mins in B array. 3C147 performed as the flux and bandpass calibrator, whereas J0414+3418 was used to be the gain calibrator.

\paragraph{Calibration~}
~All data were calibrated using the CASA pipelines \citep{casa2022} through the NRAO archive interface\footnote{\url{https://data.nrao.edu/portal}} with the versions suggested by the website depending on the dates of execution blocks (5.6.2 for blocks with BnA$\rightarrow$A array, 6.1.2 / 6.2.1 for blocks before/after 2021-Feb-27). The imaging, alignment, concatenation and self-calibration were performed using CASA 6.5.4. We inspected the calibrated data by imaging the continuum from each execution block. For the observation on 2021-Oct-17, LL polarization data from most antennas were problematic and therefore
flagged.

For the Ka band observations, each phase center was first aligned to the expected stellar position predicted by the proper motion of \diskname~ measured by GAIA DR3 \citep{gaiadr3} using the \texttt{fixvis} task, and then to the disk centers which were obtained by fitting Gaussians to the images, and with the center coordinates adopted to the common \texttt{`J2000 04h58m46.275s +29d50m36.422s'}. 
After alignment, we concatenated the observations with A array into one A-combined dataset and those with B array into one B-combined dataset for the subsequent self-calibration.

To accurately account for the large fractional bandwidth when imaging and extracting models for self-calibration, we adopted the \texttt{mtmfs} deconvolver with \texttt{nterms=2}.
We then applied a single round phase-only self-calibration to both the A-combined and B-combined data with the \texttt{gaincal} task parameters being \texttt{solint=`inf', gaintype=`G'} and \texttt{combine=`scan,spw'/`scan'} for A/B, respectively. 
The peak SNRs in continuum maps were improved by a factor of $\sim1.4$ and $\sim1.3$ for A- and B-combined data individually. 
More importantly, the disk morphology imaged by B-combined data was largely recovered from phase noise: initially the disk shows an inner bright core and some curvy bridges extending from the core, after self-calibration the disk shows the core and a recovered ring at $\sim$ 95 au consistent with that at ALMA wavelengths. 

Finally, the self-calibrated A- and B-combined data were concatenated into one, and one round phase-only self-calibration was performed to the concatenated dataset with \texttt{gaincal} task parameters being \texttt{solint=`inf', gaintype=`T', combine=`spw'}. 
For this final round, the peak SNR improved by $\sim 26\%$. Further attempts with shorter solution intervals did not have significant improvements and therefore were not applied. 
The detailed demonstration of improvements by self-calibration at each step is shown in the appendix \ref{appsec:self_cal}.
The scenario of single-round phase-only self-calibration on Ka-band data is similar to the case presented by \citet{zagaria25} for CI~Tau.

For the X band observations, we followed the same alignment procedure as for Ka band observations, and the aligned data were then concatenated. One round of phase-only self-calibration was applied, resulting in a minor improvement of $\sim 4\%$ on the peak SNR.

\subsection{ALMA Observations \& Calibration} 
\label{subsec:obs_alma}
\paragraph{ALMA Band 7~}
We used ALMA Band 7 data from project 2017.1.00470.S (PI: Leslie Looney) that provides the best available resolution data in ALMA archive of $\sim0.2''$. The observations were taken on 2018 September 18 and 19 with configuration C43-4.
The spectral windows were centered at 336.5, 338.5, 348.5 and 350.5 GHz with each spanning a bandwidth of 1.875 GHz.
Three rounds of phase-only self-calibration were performed with the solution intervals being scan length, 30 s and 15 s in order as described in \citet{harrison24}.

\paragraph{ALMA Band 6~}
~We used the self-calibrated continuum data product from the ALMA Large Program MAPS \citep{oberg21, sierra21}, where the detailed (self-)calibration procedures can be found in \citet{oberg21} and \citet{czekala21}.
Observations were taken in 2018 October and 2019 August with two correlator setup corresponding to central frequencies of 226 and 257 GHz (1.33 and 1.17 mm in wavelength). We did not combine them since they both have high SNR at at similar resolution to our Ka-band observations ($\sim 0.12''$).

\paragraph{ALMA Band 3~}
~ We combined MAPS Band 3 dataset (two correlator setup centered at 94 and 106 GHz) with the Band 3 data from project 2016.1.01042.S (PI: Claire Chandler), which has higher spatial resolution.
We included the observation on September 21, 2017\footnote{The other execution on September 09, 2017 was excluded since the ring at 95 au is significantly distorted and the disk flux scale disagrees with MAPS observations. We have tried to use the observation on Sep 21 as a self-calibration model to calibrate the observation on Sep 09: the flux was recalibrated to a consistent scale but the ring was still distorted, hence we chose not to include this execution.} 
from this higher-resolution program, two rounds of phase-only self-calibration were performed with each solution interval being \texttt{`inf'} and \texttt{`scan'} length, leading to an $\sim 15\%$ improvement on the peak SNR. Finally, we aligned it with the MAPS data and combined them (including two correlator setups in MAPS) into one for our further imaging.

\subsection{Continuum images}
\begin{figure*}[htbp]
    \centering
    \includegraphics[width=\linewidth]{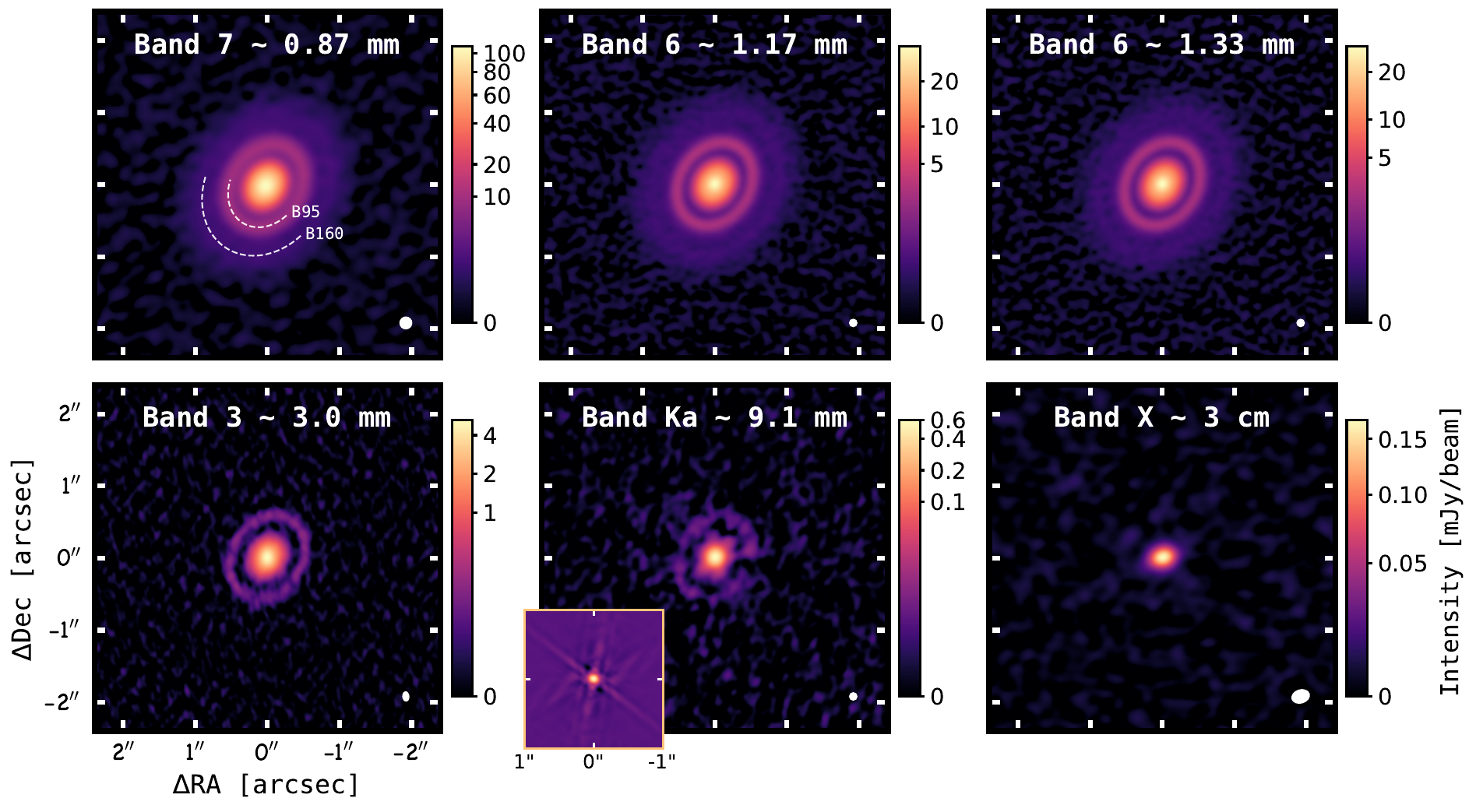}
    \caption{Continuum images of disk around MWC 480 at ALMA Band 7/6/3 and VLA Ka/X Band.
    The white dashed arcs mark the two rings and their names used in this paper.
    The color scales are in units of mJy/beam, and power or arcsinh stretches are applied to display the ring substructures and the faint outer disk.
    The inset at the lower left of Ka panel shows the point spread function (PSF) image before smoothing to a circular beam, the apparent fainter regions along the B95 ring at Ka coincide with the PSF arms.
    Beam sizes of images for each wavelength: Band 7 ($0.18''$), Band 6 ($0.12''$), Band 3 ($0.14''\times0.10''$), Ka ($0.12''$), X ($0.26''\times0.20''$).
    }
    \label{fig:multi_wave_image}
\end{figure*}

With the self-calibrated datasets as described in the above sections, we imaged MWC 480 at sets of different resolutions.
The X Band data detected the inner bright core but not the outer ring substructures, we constructed images with the \texttt{robust} parameter being 0.5 and 2.0.
For the shorter wavelengths, the lowest resolution limit is set by Band 7 which has best-achievable beam at $0.18''$ due to the lack of long baseline observations. 
The best resolution is $0.10''$, which is reachable at Bands 6 and Ka.
For the purpose of multi-wavelength analysis of dust properties, we imaged MWC 480 with circular beams of $0.18''$ (Bands 7, 6, 3 and Ka) and $0.12''$ (Bands 6, 3 and Ka).
Images with beams of $0.10''$ were created for Bands 6 and Ka to measure the width of the ring at 95 au.
To get circular beams, we adopted a combination of \texttt{robust} and \texttt{uvtaper} parameters to obtain nearly circular shapes.
Finally we use the CASA task \texttt{imsmooth} to smooth the uvtapered images to the target circular beam. 
The \texttt{multiscale} deconvolver was used for imaging, while for Band 3, Ka and X \texttt{mtmfs} (multi-term multi-frequency) deconvoler was adopted to account for their large fractional bandwidths. 
Scales of 0, 1, 3, 5 beam sizes were chosen for the deconvolving process, and we used $2\sigma$ as our clean threshold.
The continuum images with the best compromise between resolution and sensitivity at each wavelength are shown in Figure \ref{fig:multi_wave_image}.
All image properties at different resolutions are summarized in Table \ref{tab:image_properties} in the Appendix.

\section{Millimeter and Centimeter Dust Continuum Emission} 
\label{sec:obs_results}

At Bands 7 and 6, the MWC 480 disk consists of an inner bright core, a bright ring, and a fainter and more diffuse ring from inside out radially.
For convenience, we name the ring/gap features following \citet{huang18}: D73 and D144 for the two gaps; B95 and B160 for the two rings.
The numbers represent their physical radius in au and are recalculated using the updated GAIA DR3 distance.
The exact radius of these features may change at longer wavelengths, 
but the difference is negligible, so we adopt the same names for the same features at long wavelengths throughout this paper.

Figure \ref{fig:multi_wave_image} shows that the B95 ring is consistently detected from Band 7 to Ka Band.
The B160 ring is not revealed at Ka Band, but is detected in the radial profile of Band 3 (shown later) which was not seen in MAPS \citep{law21, sierra21}.
The well-resolved continuum images from Bands 7 to 3 all show a fairly axisymmetric disk morphology, while Ka emission at certain position angles along the ring is fainter than other parts.
For example, no emission is detected at the north west part that is close to disk minor axis on the Ka ring.

To quantify the level of variation along the Ka ring, we separate it azimuthally into segments of $15^\circ$ with a radial range of $0.55''-0.65''$ covering the ring. We calculate the average intensity in each segment, and the uncertainty is computed as the rms noise (Table \ref{tab:image_properties}) divided by the square root of beam numbers in the segment.
Most segments show detection levels of $3-5~\sigma$, except for those visually faint ones near north-south and east-west position angles ($0-2~\sigma$).
We notice that these segments happen to align well with the dirty PSF arms which are shown at the lower left inset of the Ka image in Figure \ref{fig:multi_wave_image}. 
Considering the high contrast (nearly a factor of 100) between the inner bright core and the B95 ring,
this apparent asymmetry is more likely due to an observational artifact that limits the dynamic range of the image even after self-calibration. MWC 480 seems to suffer from a similar dynamic range issue at VLA Ku band (2 cm), where the cleaned image shows negative arms extending from the central bright emission \citep[see Figure A.2 in][]{garufi25}.
Although the ring shows azimuthal variations that might result from low signal-to-noise ratio coupled with imaging artifacts, the radial profile is more robust.

In this section, we focus on characterizing the radial features and comparing them across different wavelengths.

\subsection{Radial intensity and visibility profiles} 
\label{subsec:result_profiles}

\begin{figure*}[htbp]
    \includegraphics[width=0.49\linewidth]{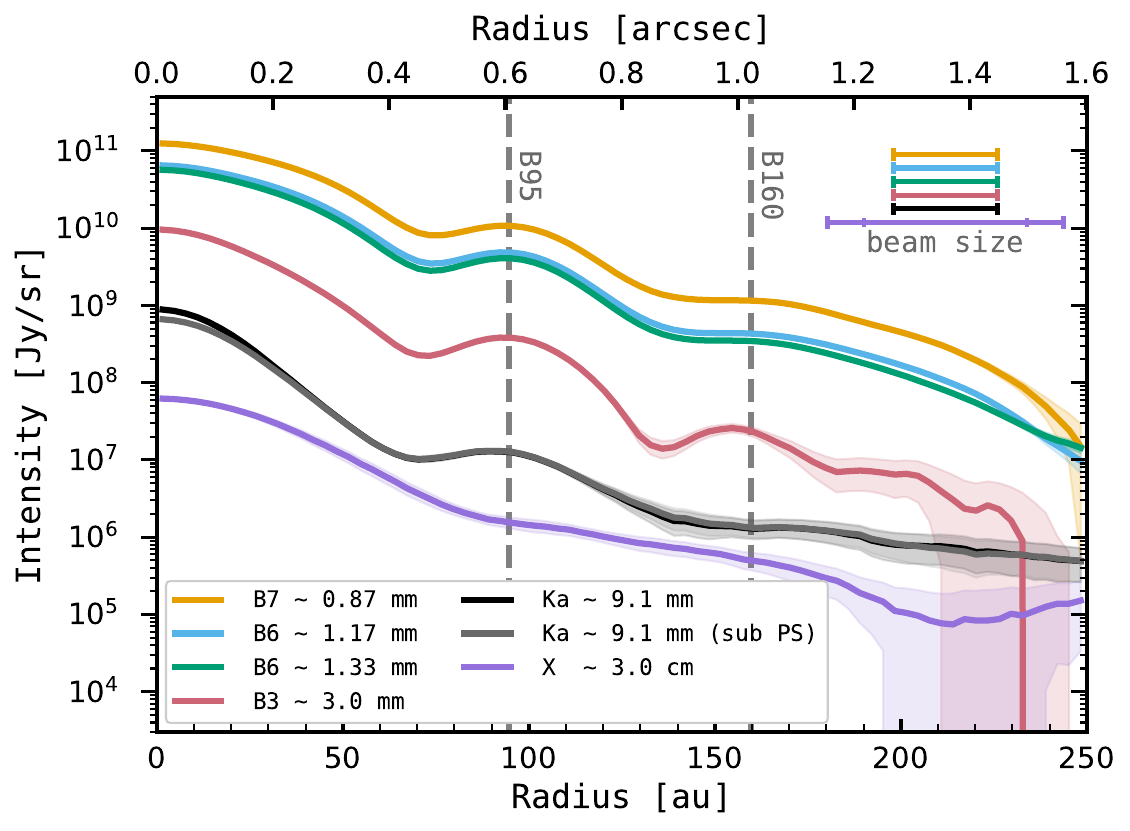}
    \includegraphics[width=0.472\linewidth]{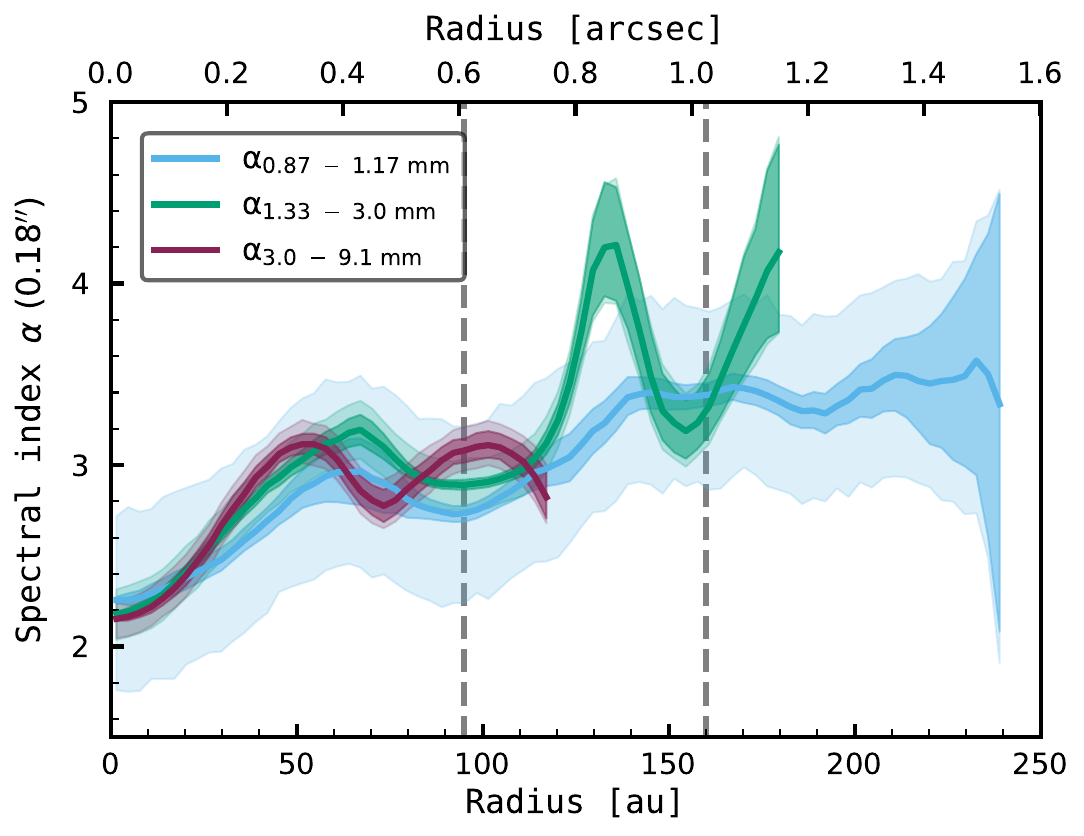}
    \caption{\textit{Left}: Deprojected azimuthally averaged radial intensity profiles, extracted from continuum images in Figure \ref{fig:multi_wave_image}.
    The shaded regions denote $1\sigma$ uncertainty.
    Two rings' positions are shown in vertical dashed gray lines.
    \textit{Right}: Spectra index profiles at resolution of $0.18''$, the deep shaded bands represent only the rms noise and the light shaded bands include the flux calibration uncertainties.
    Only radial ranges where the rms noise falls below the calibration uncertainty are displayed.
    }
    \label{fig:profiles}
\end{figure*}

\begin{figure}[htbp]
    \includegraphics[width=\linewidth]{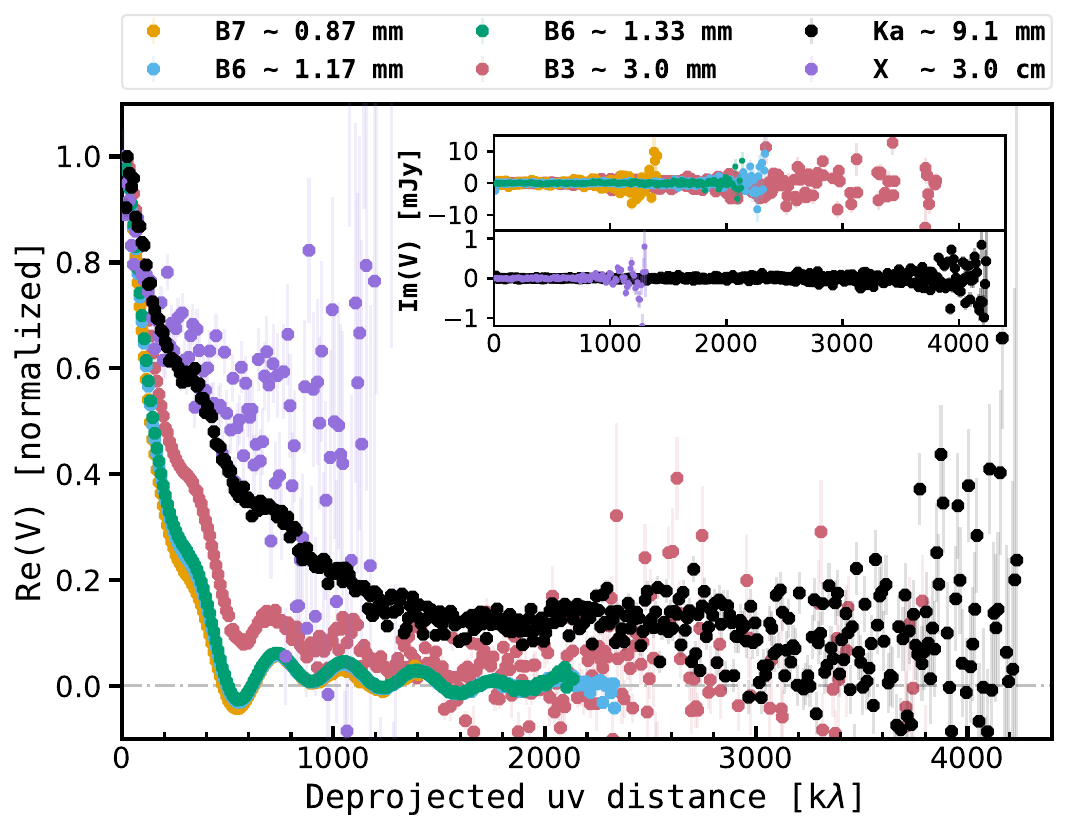}
    \caption{Deprojected visibility profiles for both real and imaginary (shown in the inset) parts.
    The real parts are normalized to the flux at shortest baselines.
    The imaginary parts for ALMA and VLA Band are shown separately in the inset, and are not normalized.
    }
    \label{fig:profiles_vis}
\end{figure}

We adopt the disk inclination and position angle to be 36.5 deg and 147.5 deg from parametric visibility modeling of Band 6 data \citep{long18} for all wavelengths. The disk center coordinates for individual images are found by fitting a Gaussian profile to the inner bright component in the image plane with the \texttt{imfit} task in CASA. 
The images are then deprojected and azimuthally averaged using \texttt{GoFish} \citep{GoFish} to get their radial profiles. The uncertainties are calculated as the standard deviation within each annulus divided by the square root of the number of beams along the annulus. The units of the profiles are converted from Jy/beam to Jy/sr. Visibilities are extracted from the self-calibrated dataset and deprojected with the corresponding geometry using \texttt{uvplot} \citep{uvplot}. Figure \ref{fig:profiles} shows the deprojected radial intensity profiles for images at resolution of $0.18''$ ($0.41''\times0.28''$ for X band) and the corresponding spectral index profiles calculated as $\alpha_{\rm mm}=\log(I_{\nu_1}/I_{\nu_2})/\log(\nu_1/\nu_2)$.
The deprojected visibility profiles for real and imaginary parts are shown in Figure \ref{fig:profiles_vis}. 

The image set with $0.18''$ beam provides a good balance between resolution and sensitivity for overall structures of the dust disk. With deep sensitivity $\sim 1.6~\mu \rm Jy/beam$, we have successfully detected the B95 ring at 9 mm.
The emission at 3 cm is spatially resolved, as indicated by its decreasing visibility profile. However, the presence or absence of the B95 ring remains uncertain due to low resolution and sensitivity, despite a possible slope turnover in the radial profile near the ring’s expected location.
For the outer part, the B160 ring identified at Band 7/6 is evident in the radial profile of the combined Band 3 data. The peak location of the B160 ring at Band 3 shifts slightly inward by $\sim 5$ au relative to Band 7/6, which is below half the beam size, blurring a robust conclusion.
This ring is not detected at 9 mm or 3 cm at the current sensitivity and resolution, and instead the profiles at these longer wavelengths show an extended tail in the outer disk.

The deprojected visibility profiles in Figure \ref{fig:profiles_vis} show a more pronounced difference among these wavelengths. From submillimeter to centimeter, the overall normalized profile is increasingly elevated or stretched toward larger uv distance. 
Since the image and visibility space have an inverse scaling property (e.g., a narrower Gaussian image corresponds to a wider Gaussian in visibility plane), the widening visibility profile of MWC 480 hints that the size of the disk emission decreases toward longer wavelength, and/or that there is a significant unresolved source contributing to the flux.

\vspace{-1pt}
\paragraph{Point source flux at 9 mm}~
Besides the overall profile, the Ka band visibility displays a plateau starting from uv distance of $\sim 1600~k\lambda$ that accounts for $\sim 12\%$ of the total flux. Since we deprojected the visibilities and shifted the disk center to the phase center, the constant offset in Ka band should be from a point-like source at the disk center given that the imaginary part is around zero. 
We used the \texttt{curve_fit} function in \texttt{scipy} \citep{scipy} to fit a constant function to the real part of the Ka deprojected visibilities beyond $1600~k\lambda$.
The point source flux is fitted to be $0.190\pm0.002\rm~mJy$.
The point source emission could be unresolved free-free and/or gyro-synchrotron emission close to the central star. 
The noisy data of Band 3 at long baseline and the lack of long baseline at X band prevent us from characterizing the point source at these two wavelengths.

\paragraph{Extended non-dust emission?}~
Recently, \citet{painter25} studied the densely sampled continuum spectra (4-360 GHz) of eight Taurus disks, including MWC 480.
They reported continuum fluxes\footnote{The reported fluxes in \citet{painter25} are measured by modeling an elliptical Gaussian to the short-baseline visibilities, essentially it is to get the total flux at the zero-baseline by extrapolating the short-baseline visibilities. Our fluxes are instead measured in the clean images with elliptical masks.} of $1.424\pm0.094$ and $1.710\pm0.099$ mJy at frequencies of 32 and 34 GHz accordingly, which are consistent with our $1.64\pm0.03$ mJy at 33 GHz (Table \ref{tab:image_properties}).
Using empirical model prescriptions, they estimated a $25^{+3}_{-3}\%$ fraction of flux at 33 GHz to be non-dust emission, which is significantly higher than our fitted point source contribution ($\sim12\%$).
This difference may point to that there might be extended non-dust emission present in MWC 480 disk at 33 GHz.
The 3 cm (X Band) observations also support the presence of resolved non-dust emission. \citet{painter25} estimated a $\sim90\%$ of emission at 3 cm to be non-dust. However, the absence of a plateau at such a level in the deprojected visilibity profile in Figure \ref{fig:profiles_vis}, indicates that the non-dust emission has a contribution from the bright inner core region.
Possible sources could be ionized gas in accretion-driven jets that are perpendicular to the dust disk plane \citep[e.g.,][]{rodriguez14, macias16} or in a photoevaporative wind \citep[e.g.,][]{pascucci12, pascucci14}.
Because of the high brightness of the inner core and low resolution and sensitivity of X Band observations, it is currently difficult to disentangle the extended non-dust emission.

\subsubsection{Disk Size versus Frequency}
\label{subsubsec:result_disk_size}

Here, we used a curve-of-growth method to calculate the disk fluxes and the effective dust disk sizes.
We applied elliptical masks with the inclination and position angle being 36.5 deg and 147.5 deg \citep{long18} with increasing size and measured the flux within the mask as a function of radius.
From this we also calculated an incremental fraction of flux at each radius.
By eye, we set a radius where the total flux within that-sized mask no longer increases monotonically and the incremental fraction of the flux begins to oscillate around zero. This cut-off radius is $1.8''$ for Band 7 and 6, $1.3''$ for Band 3, and $1.0''$ for Ka Band.
The disk fluxes were calculated within the elliptical masks with corresponding sizes, with the uncertainties being estimated as the standard deviation of the fluxes within the same elliptical masks but randomly placed at regions free of disk emission.
The disk sizes were estimated by bootstrapping radial profiles within the cut-off radius using the uncertainties extracted with \texttt{GoFish} (Section \ref{subsec:result_profiles}).

\begin{figure}[htbp]
    \centering
    \includegraphics[width=\linewidth]{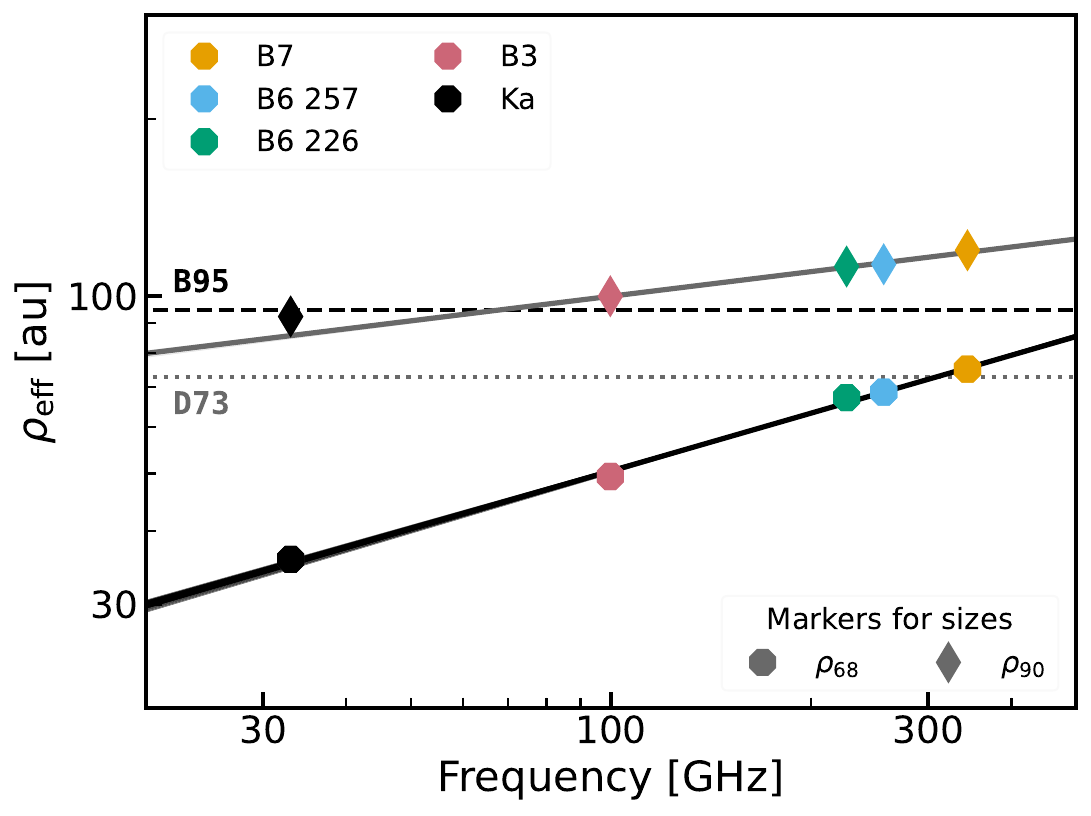}
    \caption{
    Disk sizes ($\rho_{\rm eff, 68\%}$, $\rho_{\rm eff, 90\%}$ as a function of observing frequency.
    Random draws from linear fitting to $\rho_{\rm eff, 68\%}$ and $\rho_{\rm eff, 90\%}$ are shown as solid black and gray lines.
    }
    \label{fig:disk_size}
\end{figure}

For every generated radial profile, we calculated the radius $\rho_{\rm eff,x}$ that encloses a certain fraction $x=68\%,90\%,95\%$ of the corresponding total flux $F_\nu$: $f_\nu(\rho_{\rm eff,x})=xF_\nu$ \citep{tripathi17}.
Figure \ref{fig:disk_size} shows the $\rho_{\rm eff}$ as a function of observed frequency and the curves of cumulative flux as a function of disk radius.
The disk fluxes and the $\rho_{\rm eff, 68\%}$ and $\rho_{\rm eff, 90\%}$ are summarized in Table \ref{tab:image_properties}.

Figure \ref{fig:disk_size} shows a clear trend that all measured $\rho_{\rm eff}$ increases monotonically with the observation frequency.
We fitted power-laws to $\rho_{\rm eff, 68\%}$ and $\rho_{\rm eff, 90\%}$, and we found:
\begin{equation}
    \label{eq:size_power_law}
    \begin{aligned}
        \left[\frac{\rho_{\rm eff, 68\%}}{\rm au} \right] &= (11.3\pm0.3) \times \left[\frac{\nu}{\rm GHz}\right]^{0.326\pm0.004}
        \\
        \left[\frac{\rho_{\rm eff, 90\%}}{\rm au} \right] &= (52.7\pm0.7) \times \left[\frac{\nu}{\rm GHz}\right]^{0.139\pm0.002}.
    \end{aligned}
\end{equation}
$\rho_{\rm eff, 90\%}$ displays a shallower slope than $\rho_{\rm eff, 68\%}$ with respect to frequency.
$\rho_{\rm eff, 68\%}$ for MWC 480 is similar to the measurement for UZ Tau E: $\rho_{\rm eff, 68\%}\propto \nu^{0.34\pm0.08}$ \citep{tripathi18}.
At all wavelengths, $\rho_{\rm eff, 68\%}$ fall in the inner bright disk, while $\rho_{\rm eff, 90\%}$ is located around or outside the B95 ring (see Figure \ref{fig:disk_size}).
Hence, the steeper frequency dependence of $\rho_{\rm eff, 68\%}$ may indicate that the inner smooth bright component is more drift-dominated, whereas the B95 ring, as traced by $\rho_{\rm eff, 90\%}$, is more effective at trapping dust. \citep[e.g.,][]{pinilla20}.

\subsection{Non-parametric visibility modeling}
\label{subsec:result_frank}
\begin{figure*}[htbp]
    \centering
    \includegraphics[width=0.95\linewidth]{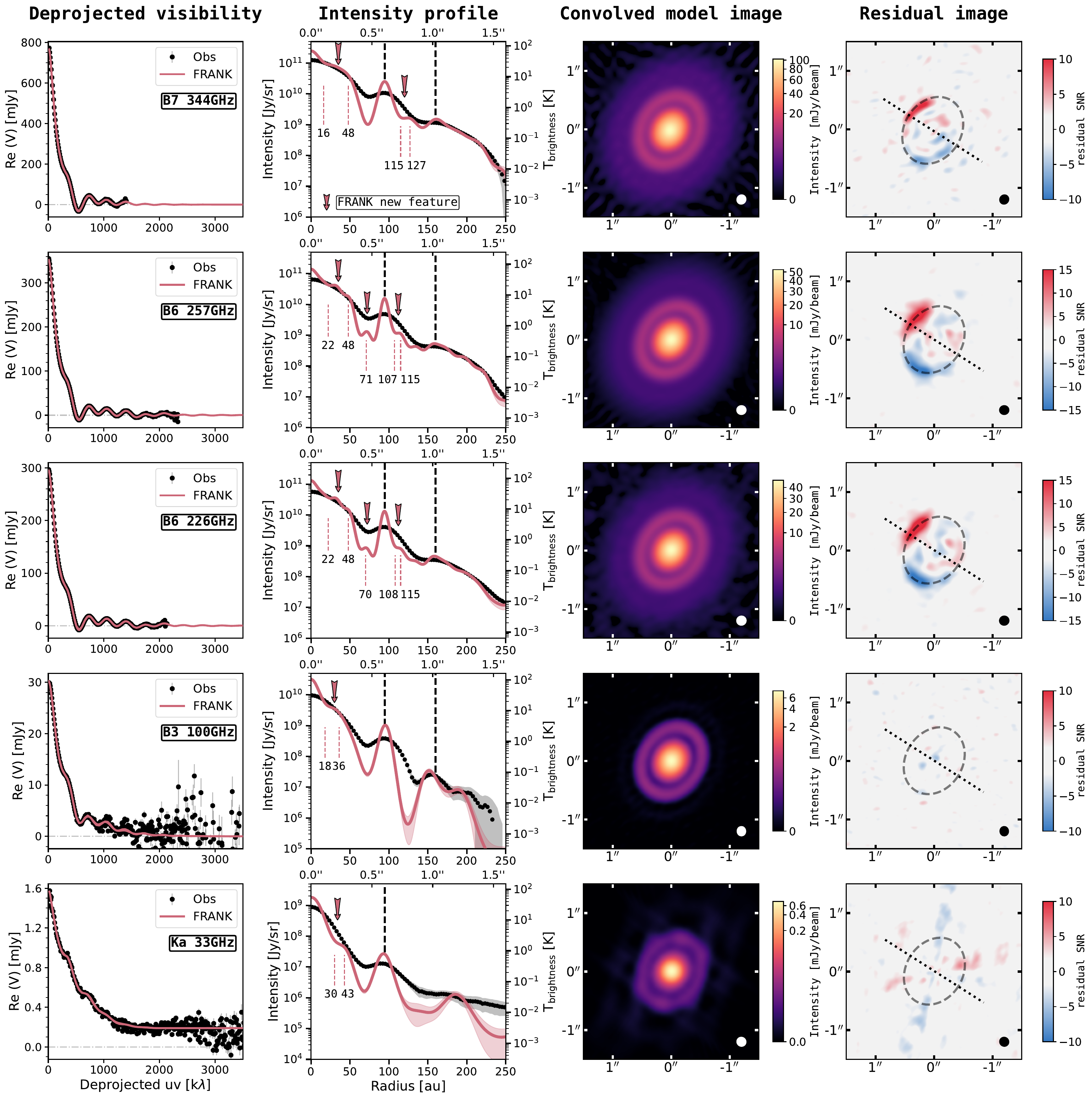}
    \caption{Results of visibility fitting using FRANK, ALMA Band 7/6/3 and VLA Ka from top to bottom.
    From left to right:
    \textit{Leftmost}: Deprojected visibility profiles. Black dots for observation and pink lines for FRANK models.
    \textit{Middle left}: Intensity profiles for observations and FRANK models, right axis also shows the converted brightness temperature.
    Dashed black lines represent the B95 and B160 rings with the radial locations from Band 7.
    Pink downward arrows denote the new structures found by FRANK fitting, and dashed pink lines label their approximate radial positions.
    \textit{Middle right}: FRANK model images convolved with observational beams of $0.18''$.
    \textit{Rightmost}: Residual images displayed in terms of signal-to-noise ratio. The B95 ring is shown as dashed gray ellipse. Disk minor axis is marked by black dotted line.
    }
    \label{fig:frank_result}
\end{figure*}

To search for any hidden substructures in the MWC 480 disk at multiple wavelengths,
We used \texttt{FRANK} \citep{frank} to fit non-parametric 1D profiles to the real part of deprojected visibilities.
The X band was not fitted due to its poor uv-coverage and low signal-to-noise ratio (Figure \ref{fig:profiles_vis}). 
Logarithmic brightness models are used in \texttt{FRANK} fitting to avoid non-physical negative brightness and largely reduce artificial oscillations in the radial profile. Conservative hyperparameters\footnote{$\alpha$ acts as a SNR threshold for the maximum baseline to which \texttt{FRANK} attempts to fit the visibilities, $w_{\rm smooth}$ sets the smoothing strength applied for estimating the power spectrum. Higher $\alpha$ imposes more strict threshold and thus fitting less noisy data, and higher $w_{\rm smooth}$ performs stronger smoothing of the power spectrum. The suggested values are: $\alpha\in[1.05, 1.30]$, $w_{\rm smooth}\in[10^{-4}, 10^{-1}]$.} $\alpha=1.3$ and $w_{\rm smooth}=10^{-1}$ were adopted to avoid overfitting the noisy visibilities at long baselines. For the Ka visibility, we subtracted the point source flux of $0.19$ mJy from the deprojected real visibilities before fitting. 
This is because \texttt{FRANK} requires the power spectrum to converge to zero beyond the maximum baselines, which would otherwise introduce strong oscillations in the brightness profiles \citep{jennings22_dsharp}.
The results of visibility, radial profile, convolved model images, and residual images are presented in Figure \ref{fig:frank_result}. The model images were constructed by inputting the corresponding visibilities into the same CASA Measurement Sets and using the same clean parameters.

\paragraph{Residuals }~
The overall axisymmetric part of the disk at all wavelengths is well fitted. In Bands 7 and 6, there are significant arc-like residuals on the B95 ring. At about $0.18''$ resolution, the northeast and southeast arcs show peaks of $\sim10/20~\sigma$ and $\sim-7/-15~\sigma$ significance at Band 7/6, respectively.
Those arc-like residuals were also identified in \citet{andrews24} using the same MAPS Band 6 data. 
In the inner disk region, residuals of $\sim5\sigma$ level are visible, while lower-level residuals can be very sensitive to the adopted disk geometry and center offsets \citep{andrews21}.
However, the arc-like features on the B95 ring are robust against these choices.
Figure \ref{fig:frank_result} also marks the disk minor axis with black dotted lines, the positive arc features extend azimuthally across disk minor axis, but are not symmetric about it.

\paragraph{Annular structures }~
The second column in Figure \ref{fig:frank_result} shows the recovered FRANK profiles - axisymmetric emission of the disk. The B95 ring in the images is clearly recovered at similar radii at all wavelengths.
The exact radius of the B160 ring appears to shift inward as the wavelength increases.
However, the B160 ring location is sensitive to the \texttt{FRANK} hyperparameters since it is in the faint-emitting region.
At Ka band, a tentative ring is recovered outside 160 au but with large uncertainties.

In addition to the two rings that appear in the images or observational radial profiles, we have also identified new annular substructures.
For the inner bright core, a plateau feature (flat slope) between 20 and 50 au is found from Band 7 to Ka, with the exact radial range depending on the wavelength (see Figure \ref{fig:frank_result}). 
This plateau at Band 6 was first reported by \citet{jennings22_taurus} using also FRANK on data from \citet{long18} which has similar uv baseline coverage as MAPS but lower SNR; later it was confirmed by \citet{yamaguchi24} using a 2D super-resolution tool \texttt{PRIISM} based on the sparse modeling technique \citep{priism}. 
With MAPS data having higher SNR at long baselines, the Band 6 plateau is actually resolved into two consecutive smaller plateaus. 
Ka plateau has the narrowest radial range, although deeper long-baseline observations are needed to measure the plateau's radial extent precisely. 
Although not necessarily physically correlated, the plateau coincides radially with rings of the $\rm CH_3CN~12-11$ and $\rm HC_3N~11-10$ lines \citep{law21}.

Faint annular features are recovered in the D73 gap and the outer edge of B95 ring. 
At two spectral setups in Band 6, we found a faint ring $\sim70$ au is located in the D73 gap, consistent with what \citet{jennings22_taurus} reported. 
At the outer edge of the B95 ring, a shoulder feature around 116 au is identified at Band 7 and 6. The shoulder at Band 6 is consistent with the inflection feature I120 reported in \citet{yamaguchi24} (considered rescaling their GAIA DR2 distance to DR3).
We visually marked the radial range of the shoulders (Figure \ref{fig:frank_result}), the Band 7 shoulder spans in (116, 128) au, which is outside the Band 6 shoulder in (108, 116) au.
However, the exact radial extent of the shoulders is likely to be affected by different baseline coverage in Bands 7 and 6.

\subsection{Ring widths across wavelength}
\label{subsec:result_ring_width}
Dust rings are often interpreted as the product of gas pressure bump, where the dust ring would have a width no larger than the width of the pressure bump. 
In this scenario, ring widths at different wavelengths probe dust trapping efficiency as they trace different grain sizes. 
We follow the steps in \citet{dullemond18} and focus on the more prominent B95 ring around MWC 480.

We first fit a Gaussian brightness profile to the B95 ring profile to obtain the observed ring width in the image plane at each wavelength:
\begin{equation}
    \label{eq:gauss}
    I^{\rm gauss}_\nu(r) = A \exp\left( -\frac{(r-r_0)^2}{2\sigma^2} \right).
\end{equation}
We use a Markov Chain Monte Carlo (MCMC) method (\texttt{emcee}, \citealt{emcee}) to find the best set of parameters $(A, r_0, \sigma)$ with 30 walkers and 2000 steps. Since the observational radial profile within one beam is not independent, we multiply the estimated uncertainty by the square root of the number of data points within one beam before calculating the likelihood.
The radial ranges used for fitting are chosen by eye where a Gaussian can well describe the shape. 
Since the convolution is in the 2D sky plane, the inclination of the disk ($i$) and the ellipticity of the beam ($b_{\rm maj},~b_{\rm min}$) will affect the extracted 1D radial profile. 
We therefore follow the procedures outlined in \citet{dullemond18} and calculated the deconvolved ring width as: 
\begin{equation}
    \label{eq:wd}
    w_d = \sqrt{\sigma^2 - \sigma^2_b}.
\end{equation}
$\sigma_b=\sqrt{b_{\rm maj}b_{\rm min}/|\cos i|}/(2\sqrt{2\ln2})$ is the effective Gaussian kernel to account for the beam convolution (see Appendix H of \citet{dullemond18}).

\begin{figure*}[htbp]
    \centering
    \includegraphics[width=0.95\linewidth]{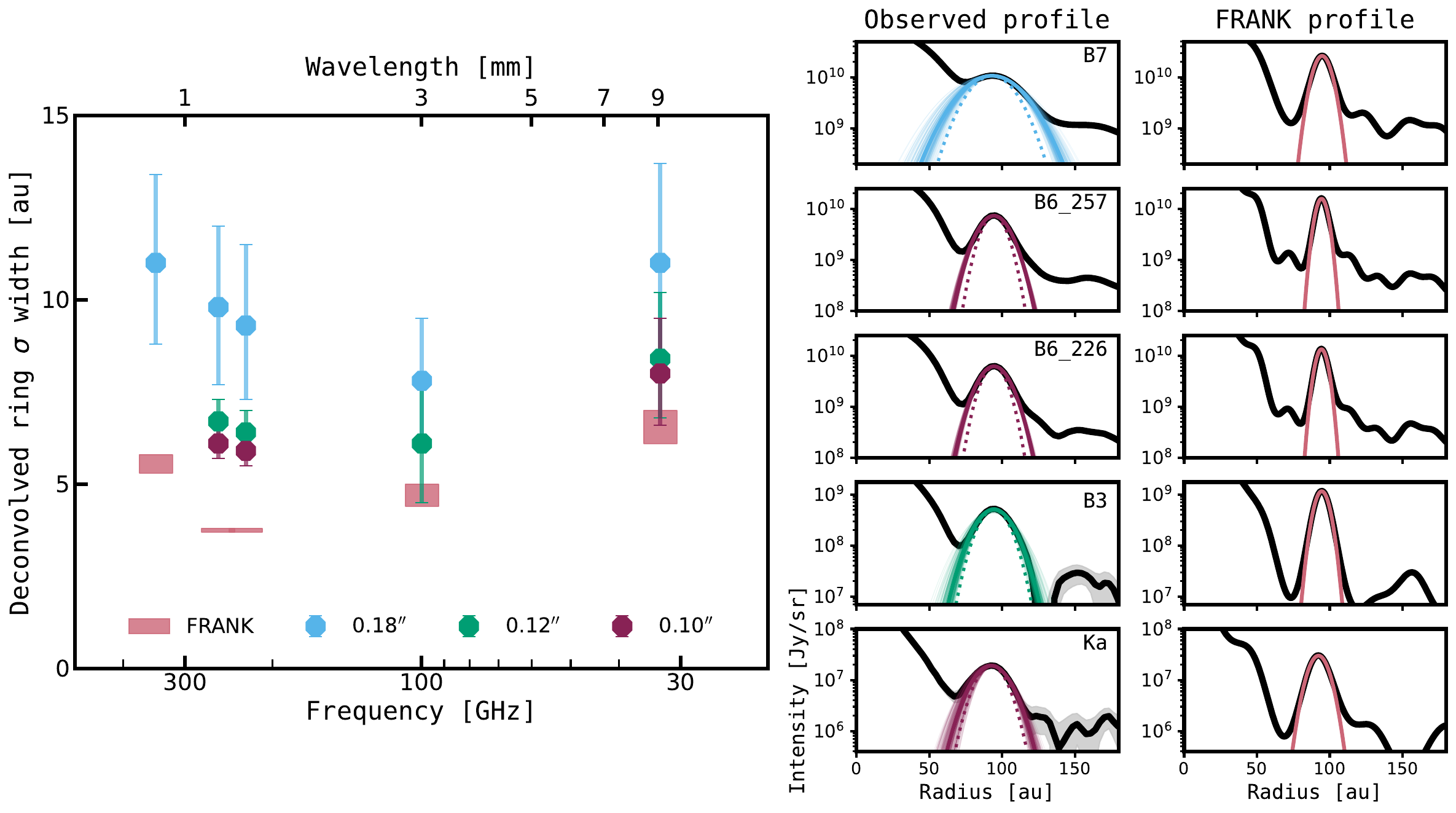}
    \caption{Measurement of deconvolved widths of the B95 ring at ALMA Band 7/6/3 and VLA Ka.
    \textit{Left}: Deconvolved ring width as a function of frequency.
    Each color represents measurement at a different resolution ($0.18''$, $0.12''$, $0.10''$) or possible range from FRANK profiles.
    \textit{Right:} Gaussian fitting to the ring intensity profiles, both from observations and FRANK models.
    For observed profiles, fitting to the profiles at the highest resolution for the corresponding frequency are shown as solid lines, with colors matching with left panel.
    The dashed parabolas denote Gaussians with the width of the beam.
    }
    \label{fig:ring_width}
\end{figure*}

This procedure was repeated for radial profiles with different resolutions of $0.18'', 0.12''~{\rm and}~0.10''$ to allow a fair comparison.
For fitting the profile at $0.12''$ at Band 3, we used image with a $0.144''\times0.1''$ beam.
This is the resolution that the combined Band 3 dataset can reach and give a much better image quality than $0.12''\times0.12''$ image from ALMA program 2016.1.01042 alone. The deprojected azimuthally averaged radial profiles have negligible difference as long as the geometric mean of the beam ($\sigma_b$) stays the same.
For reference, we also measured the B95 ring width from FRANK profiles but without further deconvolution procedure.
To account for the effect on ring profiles by FRANK hyperparameters, we performed fitting to FRANK profiles retrieved with $\alpha,w_{\rm smooth}=(1.1, 10^{-3})$ and $(1.3, 10^{-1})$ to measure the possible range of FRANK ring width.
The uv-coverage and visibility SNR may also affect the measured ring width from visibility modeling, but these effects are difficult to be quantified.
We list the results in Table \ref{tab:ring_fit} in Appendix \ref{appsec:ring_width}.
The deconvolved ring width as a function of wavelength and the Gaussian fitting to the radial profiles are shown in Figure \ref{fig:ring_width}.

The width of the deconvolved dust ring decreases when measured at higher resolution. 
One reason is that the measured ring width before deconvolution is affected by the emission at other radii, such as the inner bright disk, at lower resolution.
FRANK profiles show ring widths that are narrower than measurements from images at the highest resolution, highlighting the need for observations with even longer baselines to resolve the ring.

The resolution of $0.12''$ provides the best balance between wavelength coverage and spatial resolution.
We observe that the ring width is consistent with being constant across wavelengths within uncertainties.
This contradicts with the expectation that ring width decreases with increasing wavelength if the ring was formed by dust trapping in gas pressure bump, as longer wavelengths trace larger dust grains that are more radially concentrated.
We discuss the possible physical scenarios of a constant ring width variation in Section \ref{subsec:discuss_ring_width}.

From an observational point of view, the ring width measurement at Ka band might be limited by the image fidelity.
In Appendix \ref{appsec:self_cal} we show the improvement of the Ka image during the self-calibration.  The curvy `tentacles' extending from the inner core across the D73 gap are mostly removed, but faint residuals may persist. 
To mitigate the impact, we have selected a radial range including less inner side of the B95 ring, which results in a good fit to the ring peak and its outer emission. 
Observations with much better dynamic range are needed to robustly measure the ring width at Ka band. 
It is also important to fill in observations with high SNR and resolution at wavelengths between 3 mm and 9 mm (ALMA Bands 2 and 1), and see how the ring width in-between behaves.

\section{Modeling Dust Continuum Emission} \label{sec:model_dust}

\subsection{Method} \label{subsec:model_dust_method} 
Spatially resolved images at multiple wavelengths allow us to model the dust emission around MWC 480 as a function of radius.
Assuming a vertically isothermal disk\footnote{A good approximation for emission at millimeter which mostly come from a layer close to disk midplane.}, the emergent specific intensity from dust at frequency $\nu$ is calculated as \citep{miyake93, sierra19, carrasco-gonzalez19}
\begin{equation}
    \label{eq:intensity}
    S_\nu = B_\nu(T_{\rm dust}) [1 - \exp(-\tau_\nu/\mu) + \omega_\nu F(\tau_\nu, \omega_\nu, \mu)],
\end{equation}
where
\begin{equation}
    \label{eq:F}
    \begin{aligned}
        F(\tau_\nu, \omega_\nu, \mu) = &\frac{1}{\exp(-\sqrt{3}\epsilon_\nu\tau_\nu)(\epsilon_\nu-1) - (\epsilon_\nu+1)}
        \\
        &\times \left[\frac{1-\exp\left(-(\sqrt{3}\epsilon_\nu+1/\mu)\tau_\nu\right)}{\sqrt{3}\epsilon_\nu\mu+1} \right.
        \\
        &+ \left.\frac{\exp(-\tau_\nu/\mu) - \exp(-\sqrt{3}\epsilon_\nu\tau_\nu)}{\sqrt{3}\epsilon_\nu\mu-1} \right],
    \end{aligned}
\end{equation}
where $B_\nu(T_{\rm dust})$ is the Planck function at dust temperature $T_{\rm dust}$, $\tau_\nu=\Sigma_{\rm dust}\chi_\nu$ is the total optical depth of dust and $\chi_\nu=\kappa^{\rm abs}_\nu+\kappa^{\rm sca}_\nu$ is the extinction coefficient including the dust absorption and scattering opacities, $\omega_\nu=\kappa^{\rm sca}_\nu/\chi_\nu$ is the single-scattering albedo and $\epsilon_\nu=\sqrt{1-\omega_\nu}$, $\mu=\cos i$ accounts for the inclination of the disk.
Following \citet{carrasco-gonzalez19} and \citet{macias21},
we approximate the anisotropic scattering by using an effective scattering coefficient $\kappa^{\rm sca,eff}_\nu=(1-g_\nu)\kappa^{\rm sca}_\nu$, where $g_\nu$ is the forward scattering parameter.
In the optically thin regime, the emergent intensity is independent of albedo: $S_\nu^{\rm thin}=B_\nu(T)\kappa^{\rm abs}_\nu\Sigma_{\rm d}$, while in the optically thick regime the intensity is reduced below the Planck function: $S_\nu^{\rm thick}=B_\nu(T)[1-\omega_\nu/((\epsilon_\nu+1)(\sqrt{3}\epsilon_\nu+1)]$ \citep{sierra19}.
The direct fitted parameters in Equation \ref{eq:intensity} are the temperature, the optical depth (both absorption and scattering), and the dependence of the opacity on the frequency.

To determine the absorption and scattering opacities at each wavelength, we need to assume a grain size distribution and a set of dust properties (composition, porosity, and shape).
The size distribution is often assumed to follow a power law of number density $n(a)da\propto a^{-q}da$. 
Once the dust compositions are fixed, the dust opacity is a function of the maximum grain size $a_{\rm max}$ (opacity at millimeter wavelength is much less sensitive to the minimum grain size $a_{\rm min}$, \citealt{draine06}), and the slope $q$ which is expected to be between 2.5 and 3.5 when the maximum grain size is regulated by drift/fragmentation \citep{birnstiel12}.

\vspace{-0.1cm}
\paragraph{Dust mixture models~}
We attempted to model the multi-wavelength radial profiles with different dust mixtures.
The reference dust mixture for the DSHARP survey \citep{birnstiel18} has been frequently used by previous multi-wavelength continuum analysis. 
This dust mixture consists of 44\% refractory organics \citep{henning_stognienko96}, 36\% water ice \citep{warren_brandt08}, 17\% astronomical silicates \citep{draine03}, and 3\% troilite \citep{henning_stognienko96} in volume fraction. 
However, population synthesis models show that an absorption opacity higher than the DSHARP dust is needed to reproduce the observed spectral indices \citep{stadler22, delussu24}. 
Recent case studies, such as CI Tau \citep{zagaria25}, also have a preference for adopting the dust mixture in \citet{ricci10_taurus} but with lower porosity, in which carbonaceous materials exist as amorphous carbon \citep{zubko96}, resulting in higher opacity than the DSHARP dust with refractory organics.
Inspired by these earlier results, here we also try a different dust mixture by replacing the refractory grains in DSHARP dust with amorphous carbon produced by burning benzene proposed by \citet{zubko96}. 
The compositions of the DSHARP and Zubko dust are summarized in Table \ref{tab:dust_mixture}. 
Furthermore, we introduce different porosities of 10\%, 30\%, 50\%, 70\% and 90\% in volume to both dust mixtures.
To add porosity, we follow \citet{birnstiel18} to use the Maxwell-Garnett rule to mix vacuum (as the matrix) with the above dust mixtures.
In total, we have tested 12 models of dust mixtures.
Their opacities as a function of $a_{\max}$ at Band 6, 3, and Ka are shown in Figure \ref{fig:app_opacity} in Appendix \ref{appsec:opacity}.
Overall, Zubko mixtures have higher absorption opacties than DSHARP mixtures, and increasing porosity leads to shallower Mie resonances in the absorption opacity.

\input{table_dust_mixture}

In summary, a model of dust mixture and a set of $T_{\rm dust},~\Sigma_{\rm dust},~a_{\rm max},~q$ are needed to calculate $S_\nu$ in Equation \ref{eq:intensity}.
In this work, we chose to fix the slope $q=3.0$ instead of letting it be a free parameter\footnote{We have tested for letting $q$ being a free parameter, overall it shows an increasing trend toward larger radii similar to TW Hya \citep{macias21} and CI Tau \citep{zagaria25}. In the inner $\sim35$ au, the posterior of $q$ is rather flat between 2.5 and 3.5; the best-fit $q$ has a local maximum $\sim3.4$ at the gap and a local minimum $\sim3.0$ at the B95 ring. In the radii outside the ring, the best-fit $q$ locates between 3.5 and 4.0. In most radii, the posterior of $q$ is wide and $q=3.0$ is well located within $1\sigma$ confidence interval.}.
We used the \texttt{dsharp_opac} package \citep{birnstiel18} to calculate opacities\footnote{We used the function \texttt{get_smooth_opacities} with \texttt{extrapol=False} and \texttt{extrapolate_large_grains=False} to generate opacities.} for spherical dust grains with different compositions.

Given the inhomogeneity of resolution at different wavelengths, we follow recent studies \citep{viscardi25, zagaria25}: first we perform a spectrum fitting to lower-resolution profiles including all the wavelengths, and we then take the retrieved posterior of the maximum grain size as a prior for the higher-resolution fitting with narrower wavelength coverage. 
In the case of MWC 480, we first fitted the profiles with $0.18''$ resolution at Bands 7/6/3/Ka, then we used the posterior of the $a_{\rm max}$ to fit the profiles with $0.12''$ resolution at Bands 6/3/Ka.
This approach primarily benefits the outer disk region (beyond 120 au), as the current sensitivity of our Band 3 and Ka observations only allows for a marginal detection of the faint emission extending out to B160 in the radial profiles at a resolution of $0.18''$ (see Figure \ref{fig:profiles}), but not at $0.12''$.

For the fitting procedures, we derived posterior probability distributions and the Bayesian evidence with the nested sampling Monte Carlo algorithm MLFriends \citep{buchner16, buchner19} using the \texttt{UltraNest}\footnote{\url{https://johannesbuchner.github.io/UltraNest/}} package \citep{ultranest}.
We minimize the contamination from the non-dust emission at Ka band by subtracting the point source flux $0.19~$mJy from visibilities and then redo imaging, the Ka profile used for spectrum fitting is extracted from this new image (see this profile in Figure \ref{fig:profiles}).
For the priors of $\Sigma_{\rm dust}~\rm and~a_{\rm max}$, we adopt a log-uniform distribution: $\rm\log_{10}(\Sigma_{dust}/g\cdot cm^{-2})\in[-4,2]$, $\log_{10}(a_{\rm max}/{\rm cm})\in[-4,2]$. In the high resolution fitting, the prior of $a_{\rm max}$ instead is the posterior from low resolution fitting. 
For the prior of $T_{\rm dust}$, we followed \citet{macias21} to calculate a temperature distribution from a passively irradiated disk \citep[e.g.,][]{chiang97, dullemond01}:
\begin{equation}
    \label{eq:passive_temp}
    T(r) = \left(\frac{\phi L_*}{8\pi r^2\sigma_{\rm SB}}\right)^{1/4},
\end{equation}
where $\phi$ is the flaring angle, $L_*$ is the stellar luminosity, and $\sigma_{\rm SB}$ is the Stefan-Boltzmann constant.
The $T(r)$ profile\footnote{We refer to the Appendix E in \citet{zagaria25} for detailed derivation of the probability density function.} is calculated by varying $\phi$ uniformly within typical values 0.01-0.06 \citep[e.g.,][]{huang18} and $L_*$.
We used a stellar luminosity of $L_*=17.4L_\odot$ with an uncertainty of $0.4L_*$ \citep{vioque18}.
The log-likelihood function is:
\begin{equation}
    \label{eq:log-ll}
    \ln p(I_\nu|\Theta) = -\frac{1}{2} \sum_\nu \left[\left(\frac{I_\nu-S_\nu}{\sigma_\nu}\right)^2 + \ln(2\pi\sigma^2_\nu)\right],
\end{equation}
where $\Theta=\{T_{\rm dust},~\Sigma_{\rm dust},~a_{\rm max},~q=3.0\}$ is the parameter vector, $I_\nu$ is the azimuthally averaged brightness at frequency $\nu$ at radius $r$, $S_\nu$ is the model brightness (Equation \ref{eq:intensity}), and $\sigma^2_\nu=(\Delta I_\nu)^2+(\delta I_\nu)^2$ where $\Delta I_\nu$ is the uncertainty of $I_\nu$ extracted from continuum images and $\delta I_\nu$ is the systematic flux calibration uncertainty. 
We set $\delta I_\nu=10\%\cdot I_\nu$ for ALMA Band 7, 6 and VLA Ka Band, and $\delta I_\nu=5\%\cdot I_\nu$ for ALMA Band 3, following the suggestions by the ALMA technical handbook and the VLA observing guide\footnote{\href{https://almascience.nrao.edu/proposing/technical-handbook/}{ALMA technical handbook}; \href{https://science.nrao.edu/facilities/vla/docs/manuals/oss/performance/fdscale/}{VLA observing guide}; \href{https://science.nrao.edu/facilities/vla/docs/manuals/oss/performance/fdscale/}{VLA observational summay}}.

\subsection{Comparison of dust mixture models}
\label{subsec:model_dust_compare}

\begin{figure*}[htbp]
    \includegraphics[width=0.48\linewidth]{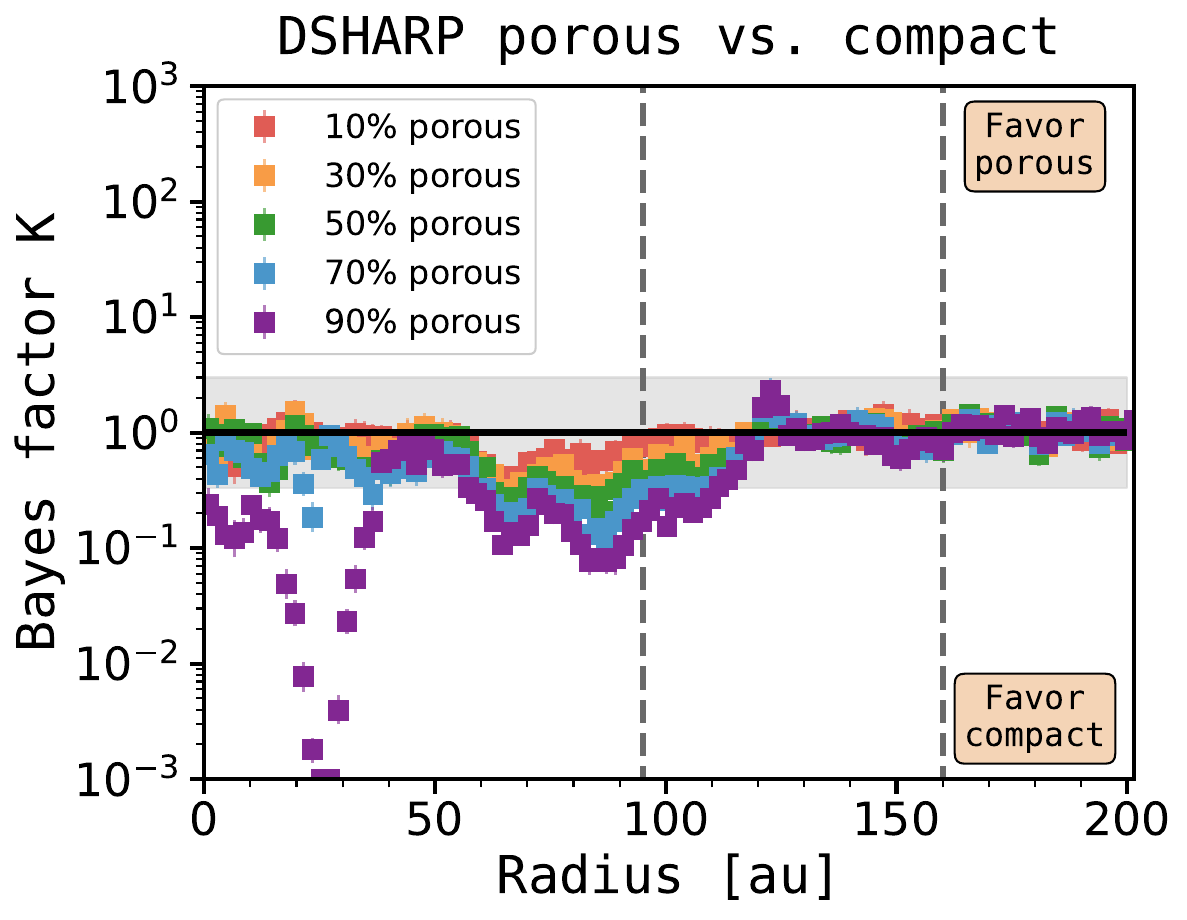}
    \includegraphics[width=0.48\linewidth]{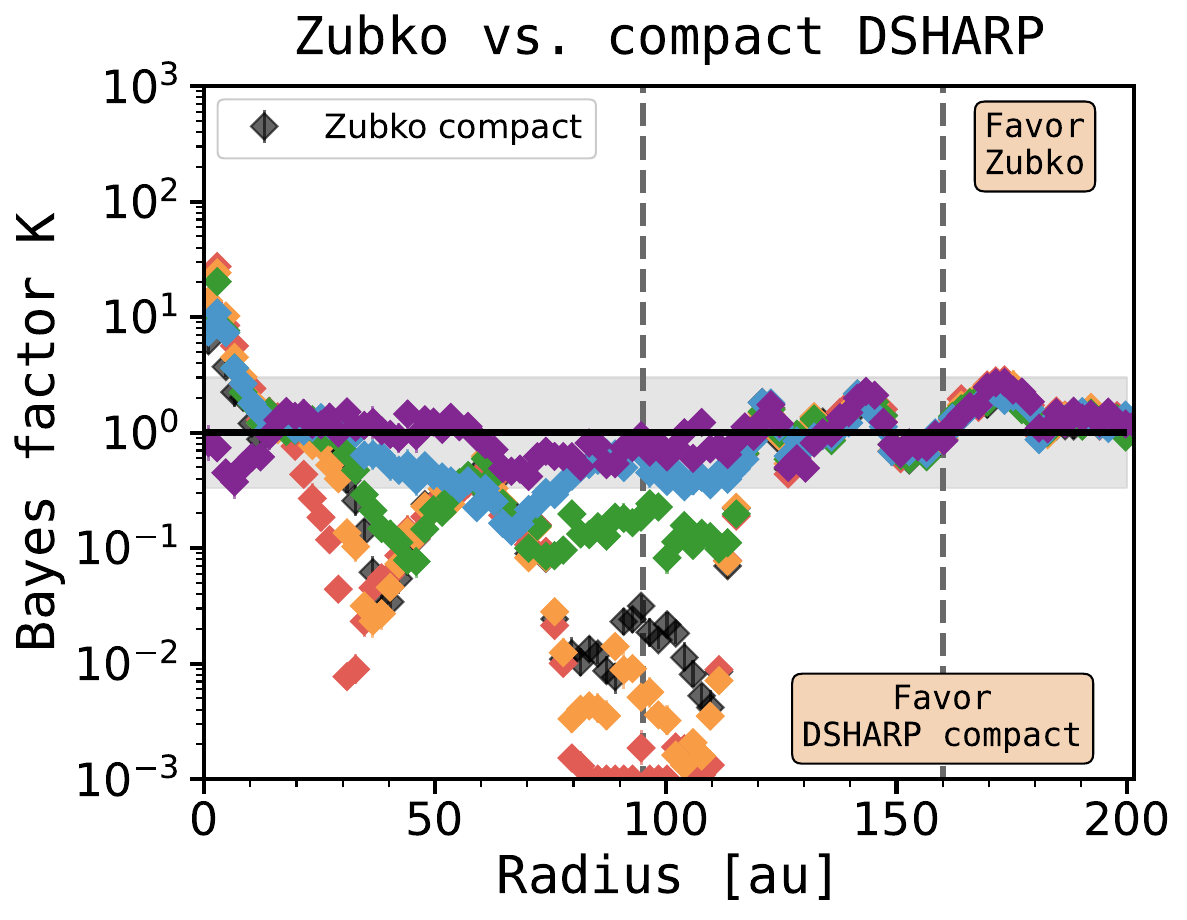}
    \caption{Bayesian comparison among dust mixture models with different carbon forms and porosities, all models are compared to the compact DSHARP mixture.
    The left and right panel represent DSHARP and Zubko mixutures with 0\%(black), 10\%(red), 30\%(orange), 50\%(green), 70\%(blue), and 90\%(purple) porosity, respectively
    Regions of $1/3<K<3$ are shaded as being ``barely worth mentioning" preference.
    }
    \label{fig:dust_compare_bayes}
\end{figure*}

We first evaluate which dust mixture models provide the best fit to our multi-wavelength data.
The nested sampling used in Section \ref{subsec:model_dust_method} provides Bayesian evidence of one model: $Z=\int L(D|\theta)\pi(\theta) {\rm d}\theta$ where $\pi(\theta)$ is the prior function and $L$ is the likelihood.
By comparing the evidence of different models, we can calculate the Bayesian factor $K=Z_{\rm dust1}/Z_{\rm dust2}$ to assess which model is preferred, assuming that two models have the same prior probability.

Figure \ref{fig:dust_compare_bayes} shows $K$ as a function of radius, where $K$ is calculated by comparing each dust mixture to the compact DSHARP mixture.
The left panel is for DSHARP mixtures and the right panel is for Zubko mixtures.
Following the scale proposed by \citet{jeffreys1939}, the ``barely worth mentioning" evidence in favor of either mixture is shaded gray ($1/3<K<3$).
$K>3$ means that the dust mixture is preferred more than the compact DSHARP.
For regions outside 120 au, there is no significant preference among the 12 dust mixtures, likely due to the low S/N of the radial profiles in Band 3 and Ka.
For  DSHARP mixtures, compact DSHARP is preferred within 120 au.
The 10-40 au region (overlap with the plateau feature in Section \ref{subsec:result_frank}) and the B95 ring disfavor highly porous (90\%) DSHARP mixture.
For Zubko mixtures within 120 au, porosity $>50\%$ is favored more than compact Zubko.
Overall, the 90\% porous Zubko mixture provides an similarly good fit as the compact DSHARP mixture, except for the inner 10 au where Zubko mixture with $\leq70\%$ porosity is more favored instead.

\subsection{Radial profiles of dust properties} \label{subsec:model_dust_profiles}

\begin{figure*}[htbp]
    \centering
    \includegraphics[width=0.96\linewidth]{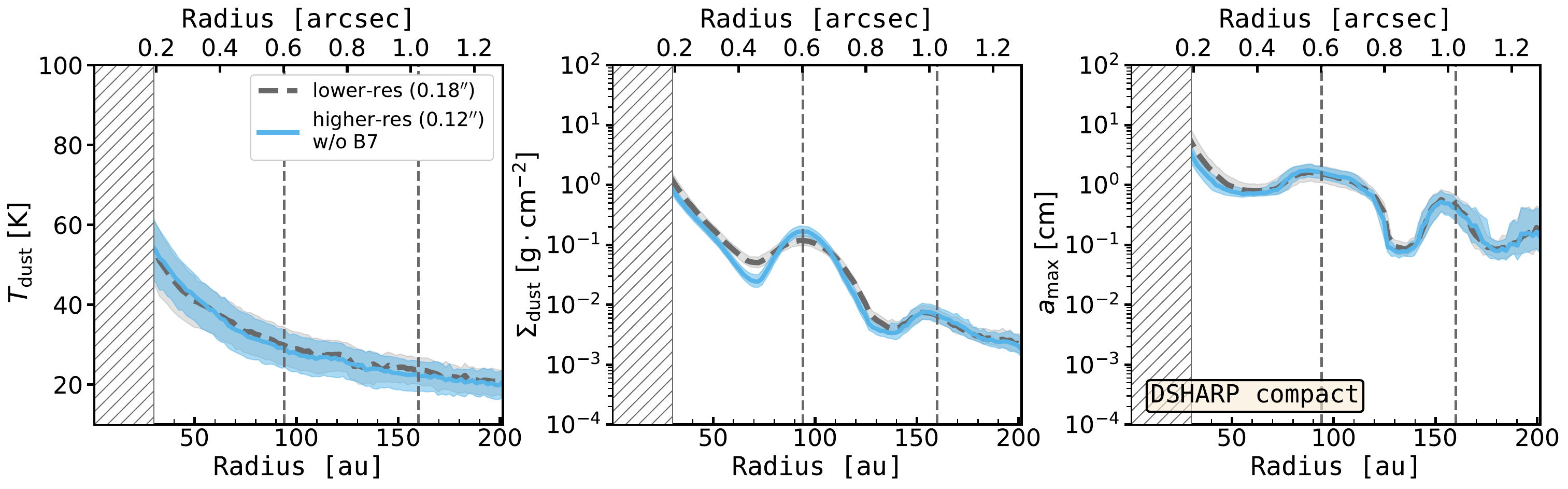} \\
    \includegraphics[width=0.96\linewidth]{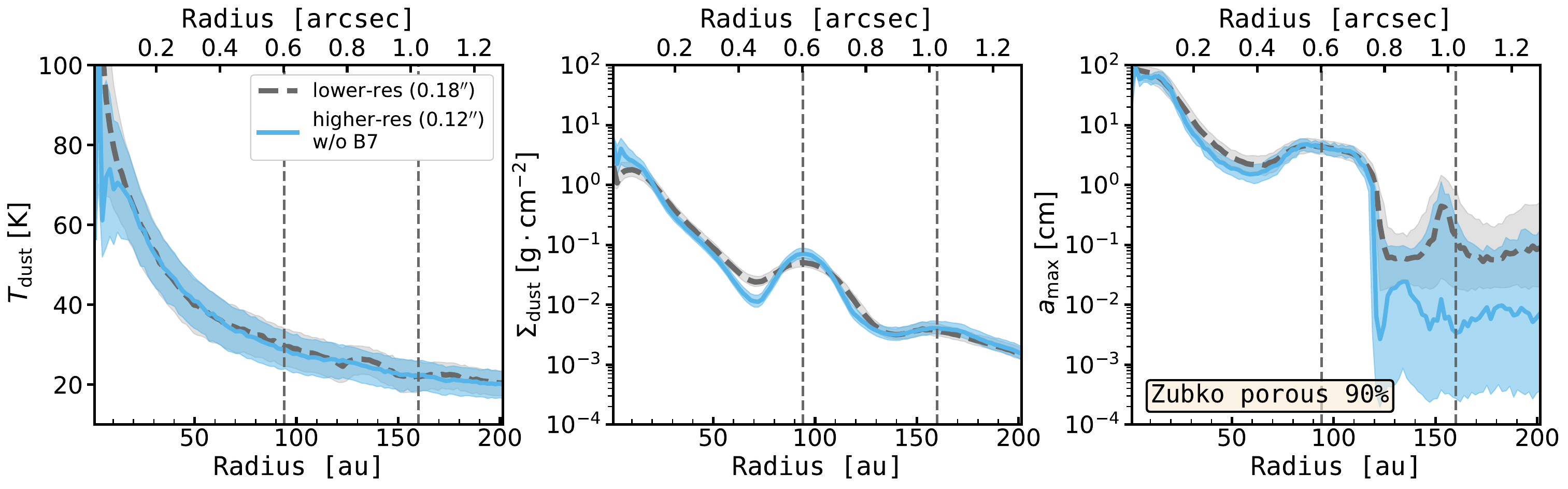} 
    \caption{Posterior distributions of dust properties for compact DSHARP (upper row) and 90\% porous Zubko dust (lower row), showing dust temperature (left column), surface density (middle column), and maximum grain size (right column). 
    Dashed gray lines denote the posteriors from lower resolution fitting ($0.18''$), and solid blue lines are for higher resolution fitting ($0.12''$).
    The shaded regions display $1\sigma$ uncertainty.
    The inner 30 au for DSHARP dust are hatched due to the presence of two branch solutions, which is better displayed in probability colormap in Figure \ref{fig:dsharp_dust_inner_disk}.
    B95 and B160 rings are shown as vertical dashed gray lines.
    }
    \label{fig:dust_props}
\end{figure*}

We adopt compact DSHARP and 90\% porous Zubko as fiducial dust mixture models, as they provide fits of similar quality.
Figure \ref{fig:dust_props} shows the results of $T_{\rm dust},~\Sigma_{\rm dust},~{\rm and}~a_{\rm max}$ retrieved for these two mixtures.
The dashed gray lines and their shaded region display the median and $1\sigma$ distribution ($16-84\%$) of the profiles from fitting at a lower resolution of $0.18''$.
The solid blue lines and the light-blue-shaded regions are for profiles with a higher resolution of $0.12''$ (with $a_{\rm max}$ from $0.18''$ fitting as priors). 

\begin{figure*}[htbp]
    \centering
    \includegraphics[width=1\linewidth]{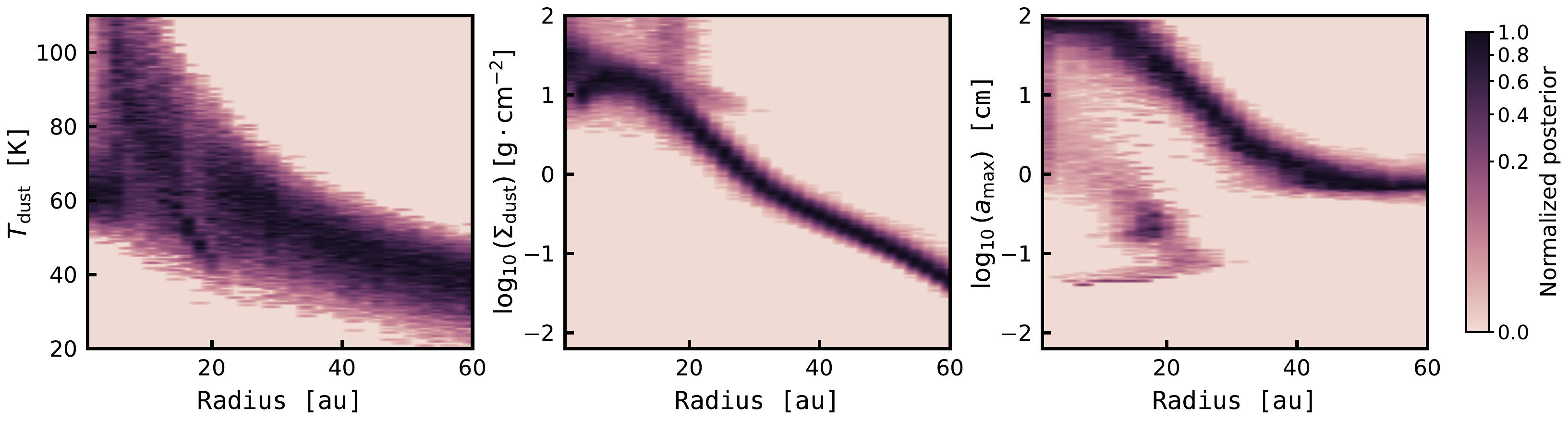}
    \caption{Normalized posterior probability colormap for compact DSHARP dust in the inner bright core.
    Color scale has been stretched to hightlight solutions with smaller maximum grain sizes.
    }
    \label{fig:dsharp_dust_inner_disk}
\end{figure*}

The comparison of the observed and modeled brightness profiles is shown in Figure \ref{fig:app_obs_model_profile_dsharp} (compact DSHARP) and \ref{fig:app_obs_model_profile_zubko} (90\% porous Zubko) in Appendix \ref{appsec:obs_model_profile}.
Multi-wavelength emission at most radii is reproduced well.
However, both mixtures fail to reproduce Ka emission at the D73 gap and the inner side of the B95 ring, and also at the regions outside 120 au.
Emission in the gap region could be due to the residual phase noise in Ka image, or the assumption of a single dust population with a power-law size distribution does not hold.
The outer part is likely due to the low SNR of the Ka band data.

The posterior of dust temperature is mostly set by the prior from a passively irradiated disk, which is similar to a few previous cases of different disks (e.g., \citealt{macias21}; \citealt{zagaria25}).
In dust surface density profiles, both dust mixtures show prominent peaks on the B95 ring, but only small bumps on the B160 ring.
The overall dust surface density decreases from the inner disk to the outermost regions, with a drop of two orders of magnitude for both dust.

The inferred maximum grain sizes exhibit local maxima on both rings, indicating local enhancement of dust growth in ring substructures.
At large distances ($>$ 120 au), compact DSHARP dust shows $a_{\rm max}$ being a few millimeters, while porous Zubko displays a broad range below compact DSHARP.
This is because the 90\% porosity leads to much shallower Mie resonances in absorption opacity of Zubko mixture.
In the B95 ring, $a_{\rm max}$ reaches centimeter sizes for two mixtures and shows a shallow variation on the ring.
Between 30 and 70 au, $a_{\rm max}$ of 90\% porous Zubko (range from $\sim$7.1 - 1.9 cm) is around 2.5 times larger than compact DSHARP (range from $\sim$2.9 - 0.8 cm) at most radii.
Within 30 au, $a_{\rm max}$ for the 90\% porous Zubko reaches $\sim$ 50 cm toward disk center, while DSHARP mixture shows solutions of large and small $a_{\rm max}$ (Figure \ref{fig:dsharp_dust_inner_disk}).
The large solution of $a_{\rm max}$ goes to high-end of allowed values (100 cm)toward disk center, while the small solutions range from sub-mm to cm.

\vspace{-0.2cm}
\paragraph{Flat grain sizes on the B95 ring }~
Previous multi-wavelength analysis of MWC 480 using lower resolution data at Band 6 (257 and 226 GHz) and Band 3 \citep{sierra21} has found the $a_{\rm max}$ is nearly constant within 120 au.
They have a fixed dust temperature profile, and used a grain size distribution slope $q=2.5$, which causes less radial variations\footnote{We have done the same exercise and obtained similar results.}.
Our results with $q=3.0$ show a radial modulation of grain size.
However, the flat $a_{\rm max}$ across the B95 ring seems to be robust: $\sim1.5$ cm for compact DSHARP and $\sim4$ cm for 90\% porous Zubko.
Such a feature indicates that the dust growth is limited by collisional fragmentation (e.g., \citealt{jiang24}), which further could be used to infer the disk turbulence level.

\vspace{0.2cm}
Figure \ref{fig:dsharp_dust_inner_disk} displays that the inner 30 au has a small grain solution of $\sim500\mu{\rm m}$ - 1 mm for compact DSHARP mixture, which is consistent with the results in \citet{sierra21}, despite the difference caused by the different choices of $q$.
We show our reproduced results in Appendix \ref{appsec:reproduce_maps} following the approach in \citet{sierra21}.
This solution is interesting because it is close to the constraint from polarization observations.
By modeling the dust polarization patterns at 0.87 mm and 3 mm in the MWC 480 disk, \citet{harrison24} have found $a_{\rm max}\sim200\mu\rm m$ with high optical depth, or two dust populations with $a_{\rm max}\sim~490\mu{\rm m}$ near the midplane and $a_{\rm max}\sim~140\mu{\rm m}$ elevated in the disk atmosphere can reproduce the polarization patterns.
Though beyond the scope of this work, an analysis of dust in MWC 480 combining multi-wavelength continuum and polarization observations is promising to break the degeneracy.

\begin{figure*}[htbp]
    \centering
    \includegraphics[width=1\linewidth]{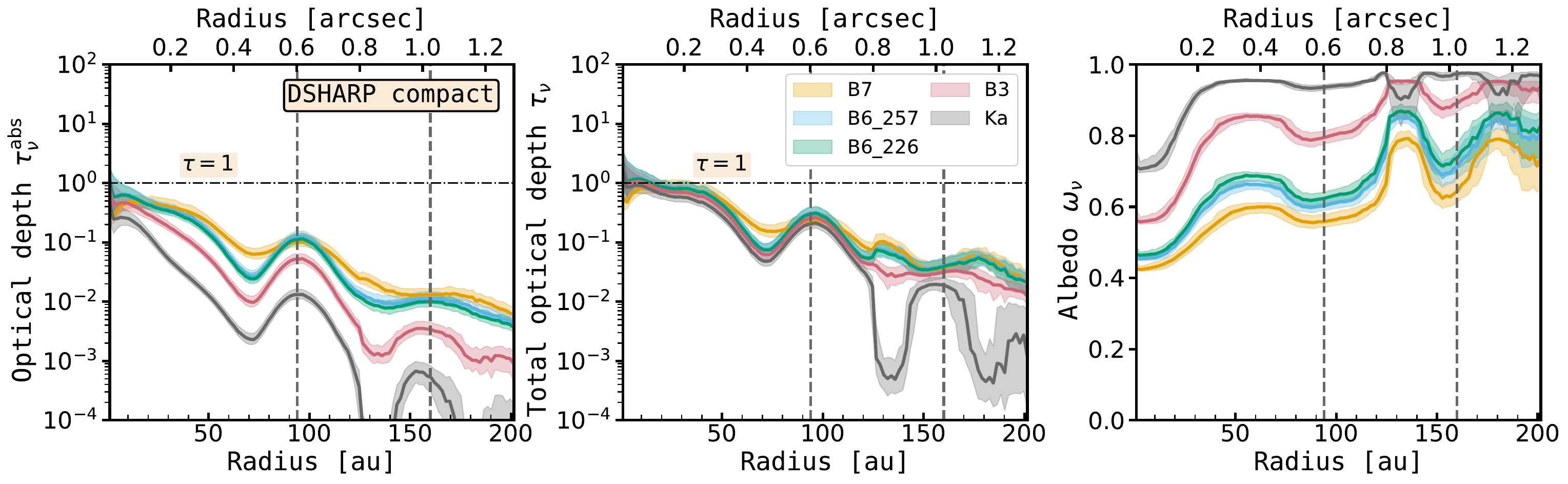}\\
    \includegraphics[width=1\linewidth]{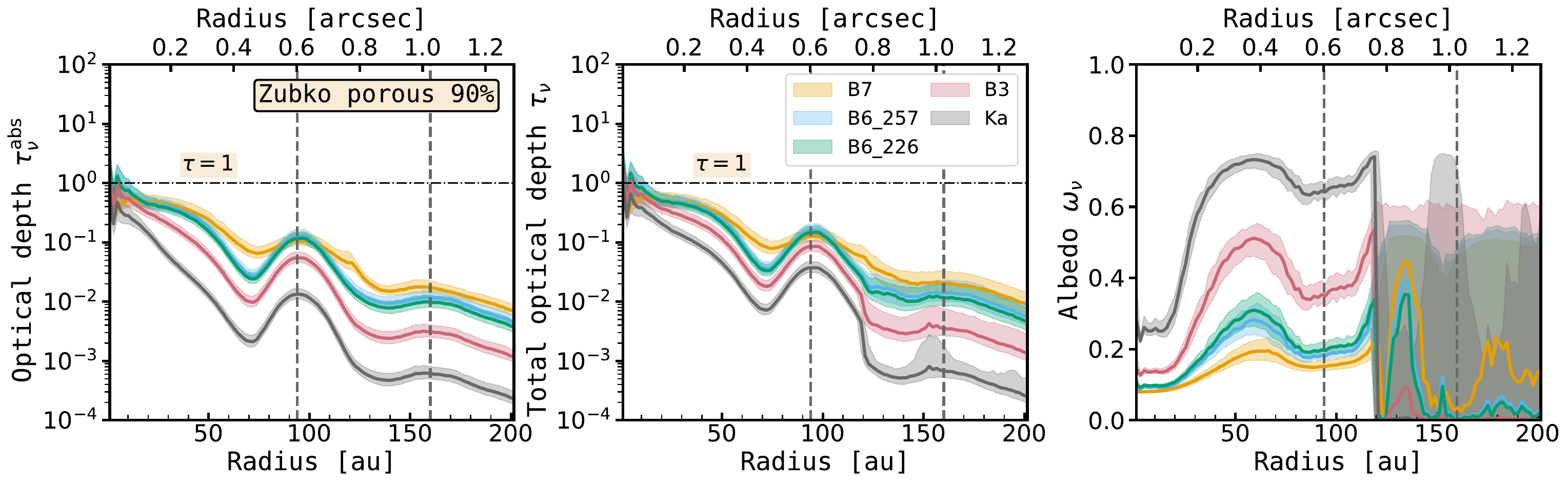} 
    \caption{Posterior distributions of dust absorption optical depth (left), total optical depth (middle), and albedo (right) for compact DSHARP (upper row) and Zubko with 90\% porosity (lower panel) dust mixture models.
    The results for Band 7 is from fitting at $0.18''$, while others are at $0.12''$.
    $\tau=1$ is marked as horizontal dash-dotted line in the left two panels.
    }
    \label{fig:dust_opt_depth}
\end{figure*}

\subsection{Dust optical depth and albedo}
\label{subsec:model_dust_optical_depth}
Figure \ref{fig:dust_opt_depth} shows the median and 1$\sigma$ uncertainties of dust optical depths and dust albedo at each wavelength calculated from the posterior distributions in Section \ref{subsec:model_dust_profiles}.
The solid lines display the results of fitting to profiles at $0.12''$ ($0.18''$ for Band 7).
For compact DSHARP dust in regions within 30 au, we show the optical properties of the large-grain solution.
Compact DSHARP dust has similar absorption optical depth as porous Zubko, but higher total optical depth due to its higher dust albedo.
The total optical depth $\tau_\nu$ overall decreases toward the outer disk and longer wavelengths, and peaks locally around the dust rings.
The region within 10 au is optically thick across wavelengths of 0.9 - 9 mm for the compact DSHARP dust, but only from 0.9 - 1.3 mm for in the case of 90\% porous Zubko dust.
Both dust mixtures show that the B95 ring is optically thin ($\tau_\nu<1$) at all wavelengths.
The dust albedo $\omega_\nu$ shows a positive dependence on the wavelengths, and displays local minimum at the B95 ring due to the larger $a_{\rm max}$.

\section{Discussion} 
\label{sec:discussion}

\subsection{Optically thick inner disk?}
\label{subsec:discuss_inner_disk}

In Section \ref{subsec:result_frank}, our visibility modeling using \texttt{FRANK} shows that the brightness profiles exhibit a plateau (or flat slope) from $\sim$20 - 50 au, with the exact range depending on the wavelength. Within 20 au, the brightness increases toward the disk center at a steeper slope than on the plateau.
In the case of compact DSHARP mixture, Figure \ref{fig:dust_opt_depth} shows that the shallower slope on the intensity plateau could be attributed to the transition from optically thin to optically thick towards the inner radius.
The emergent intensity $S_\nu(r)$ scales as $S^{\rm thin}_\nu(r)=B_\nu[T(r)]\kappa^{\rm abs}_\nu(r)\Sigma_{\rm d}(r)$ in the optically thin region, while it saturates to a reduced Planck function in the inner optically thick region: $S^{\rm thick}_\nu=B_\nu[T(r)][1-\omega_\nu(r)/((\epsilon_\nu(r)+1)(\sqrt{3}\epsilon_\nu(r)+1)]$.
Therefore, the plateau appears because $S_\nu$ transits from optically thin to thick toward the inner radius and becomes insensitive to the increasing dust surface density.
The steeper slope within 20 au is likely due to the steeper increase in dust temperature because of the closer distance to the central star.
The optically thick inner core might also hide dust substructures, as the case of MP Mus (PDS 66) which shows smooth brightness profile at 1.3 mm but displays a cavity and a ring at 3 mm \citep{ribas25}. 
However, in the case of 90\% porous Zubko mixture, the total optical depth is only close to 1 between 0.9 and 1.3 mm and thinner at longer wavelengths at the radial ranges of the plateau.
Hence, the optical depth cause for the apparent plateaus at all wavelengths is inconclusive here and further investigation is needed.

\subsection{Dust masses}
\label{subsec:discuss_dust_mass}

\begin{figure}[htbp]
    \centering
    \includegraphics[width=1.0\linewidth]{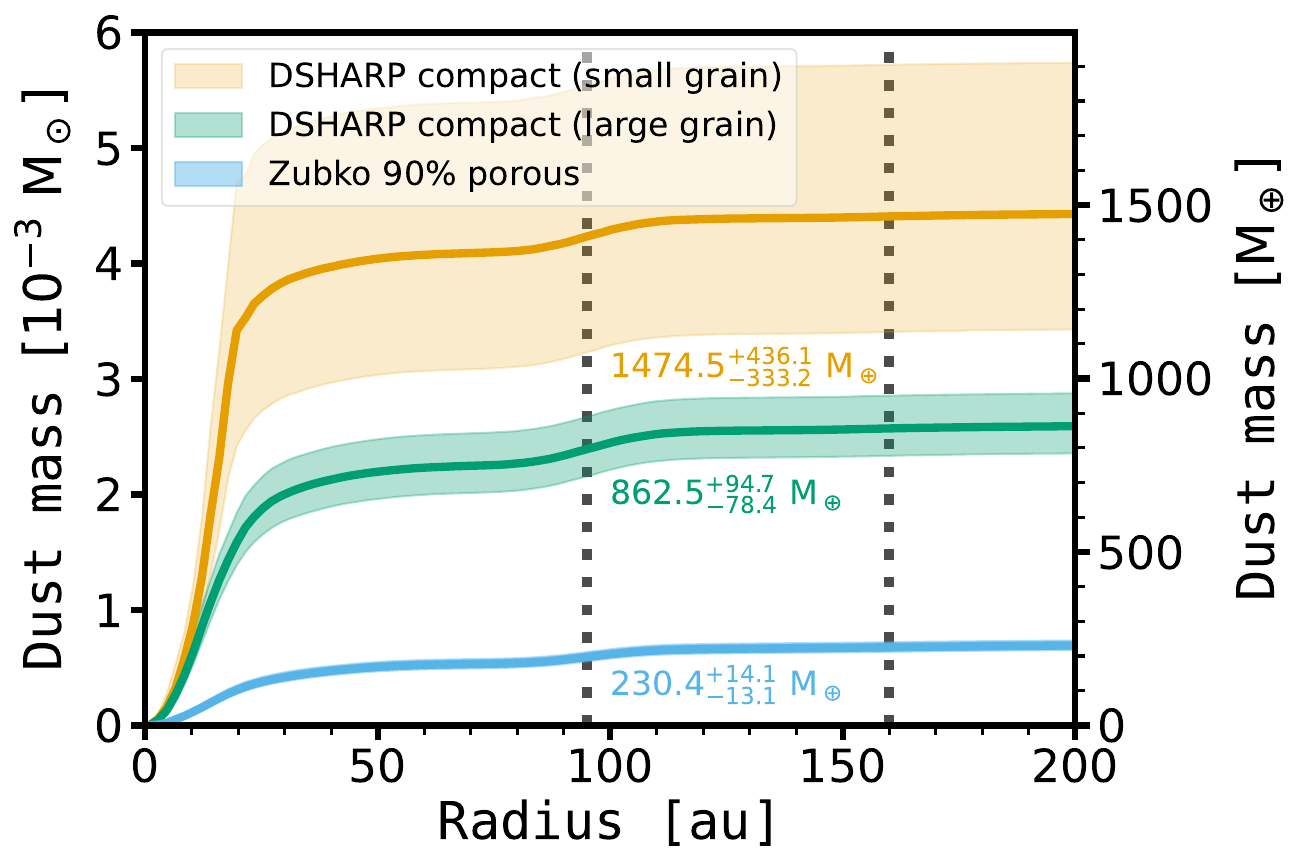}
    \caption{
    Cumulative dust mass integrated from disk center outward, for compact DSHARP (small and large grain solution separated) and 90\% porous Zubko dust mixtures.
    The numbers indicate the integrated dust masses within 200 au in unit of Earth mass with the same color as their corresponding mixtures.
    Vertical dotted gray lines denote the B95 and B160 rings.
    }
    \label{fig:cumu_dust_mass}
\end{figure}

The dust masses in disks are important to evaluate the ability to form planets.
Using the density profiles inferred in Section \ref{subsec:model_dust_profiles}, we compute dust masses as: $M_{\rm d}=\int_{r_{\rm in}}^{r_{\rm out}}\Sigma_{\rm d}(r)2\pi r {\rm d}r$.
We integrate dust masses from the disk center outward, the cumulative masses as a function of radius are shown in Figure \ref{fig:cumu_dust_mass}.
For compact DSHARP mixture, the total dust masses within 200 au are $2.6^{+0.3}_{-0.2}\times10^{-3}\rm~M_\odot$ ($860^{+95}_{-78}\rm~M_\oplus$) for the large grain-solution and $4.4^{+1.3}_{-1.0}\times10^{-3}\rm~M_\odot$ ($1500^{+440}_{-330}\rm~M_\oplus$) for the small-grain solution in the inner 30 au.
The 90\% porous Zubko gives a total mass of $6.9^{+0.4}_{-0.4}\times10^{-4}\rm~M_\odot$ ($230^{+14}_{-13}\rm~M_\oplus$).
The cumulative mass plot also shows that the inner bright core hosts most of the dust masses: the mass within 70 au occupies $\sim$93\%, 85\% and 77\% of total dust mass for DSHARP small-grain solution, large-grain solution and 90\% porous Zubko, respectively.

Before this work, \citet{sierra21} have performed resolved multi-wavelength analysis for the MWC 480 disk with DSHARP dust (from Band 6 to Band 3, fitting $\Sigma_{\rm d}, {\rm and}~a_{\rm max}$ while fixing $T_{\rm d}$, adopting $q=2.5$).  
Integrating the surface densities to $\rho_{\rm eff, 95\%}$ at Band 3 (110 au), they reported dust masses (rescaled using the GAIA DR3 distance) $1.1^{+1.7}_{-0.3}\times 10^{-3}\rm~M_\odot$ for the large-grain solution and $\sim4.2\times 10^{-3}\rm~M_\odot$ for the small-grain solution.
Within $1\sigma$ uncertainties, our calculations agree with \citet{sierra21}, despite different approaches to fit multi-wavelength radial profiles.
Earlier works using radiative transfer models have reported constraints on dust masses in the MWC 480 disk to be $\sim 3.2\times 10^{-3}M_\odot$ \citep{pietu06}, $\sim 2.3\times 10^{-3}M_\odot$ \citep{guilloteau11}, and and $1.6^{+0.5}_{-0.4}\times 10^{-3}M_\odot$ \citep{liu19}.
Our estimated mass using compact DSHARP with large grain-solution is closest to these measurements, within a factor of 0.6 - 1.2.

Taking the radial range between 70 and 120 au, we estimate the dust masses accumulated in B95 ring: $100^{+5}_{-5}~M_\oplus$ for compact DSHARP and $43^{+2}_{-2}~M_\oplus$ for 90\% porous Zubko opacity.
Hence, the dust in the B95 ring might be capable of assembling the core of a giant planet ($\sim10\rm~M_\oplus$), given high efficiency of pebble accretion 
\citep[$\sim30\%$ for Zubko while lower for DSHARP, see][]{drazkowska23}.
The inner disk within the B95 ring which hosts hundreds of $\rm M_\oplus$ of dust, can be an efficient place for forming gas giants.

\subsection{Dust composition and porosity}
\label{subsec:discuss_dust_compo}

\begin{figure*}[htbp]
    \includegraphics[width=0.48\linewidth]{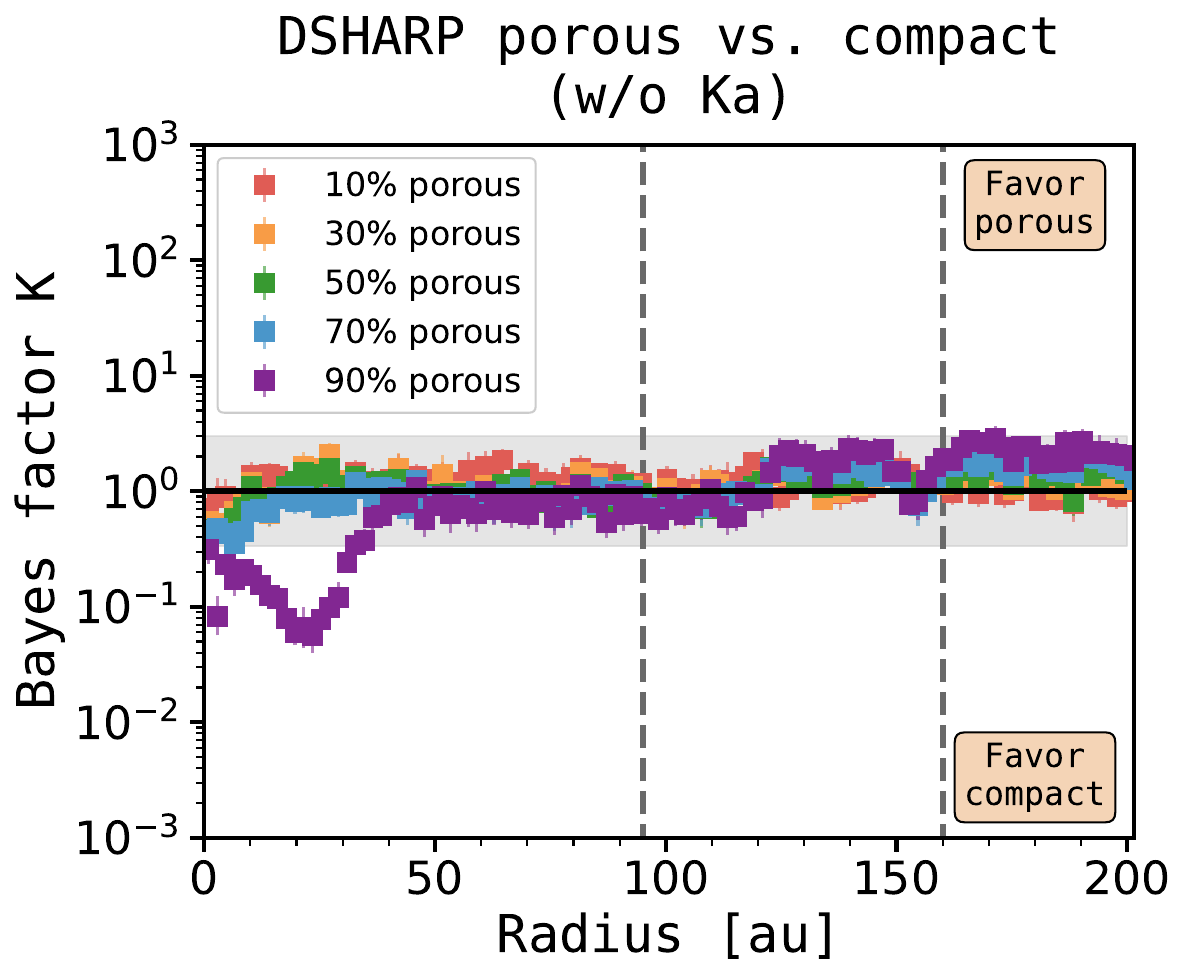}
    \includegraphics[width=0.48\linewidth]{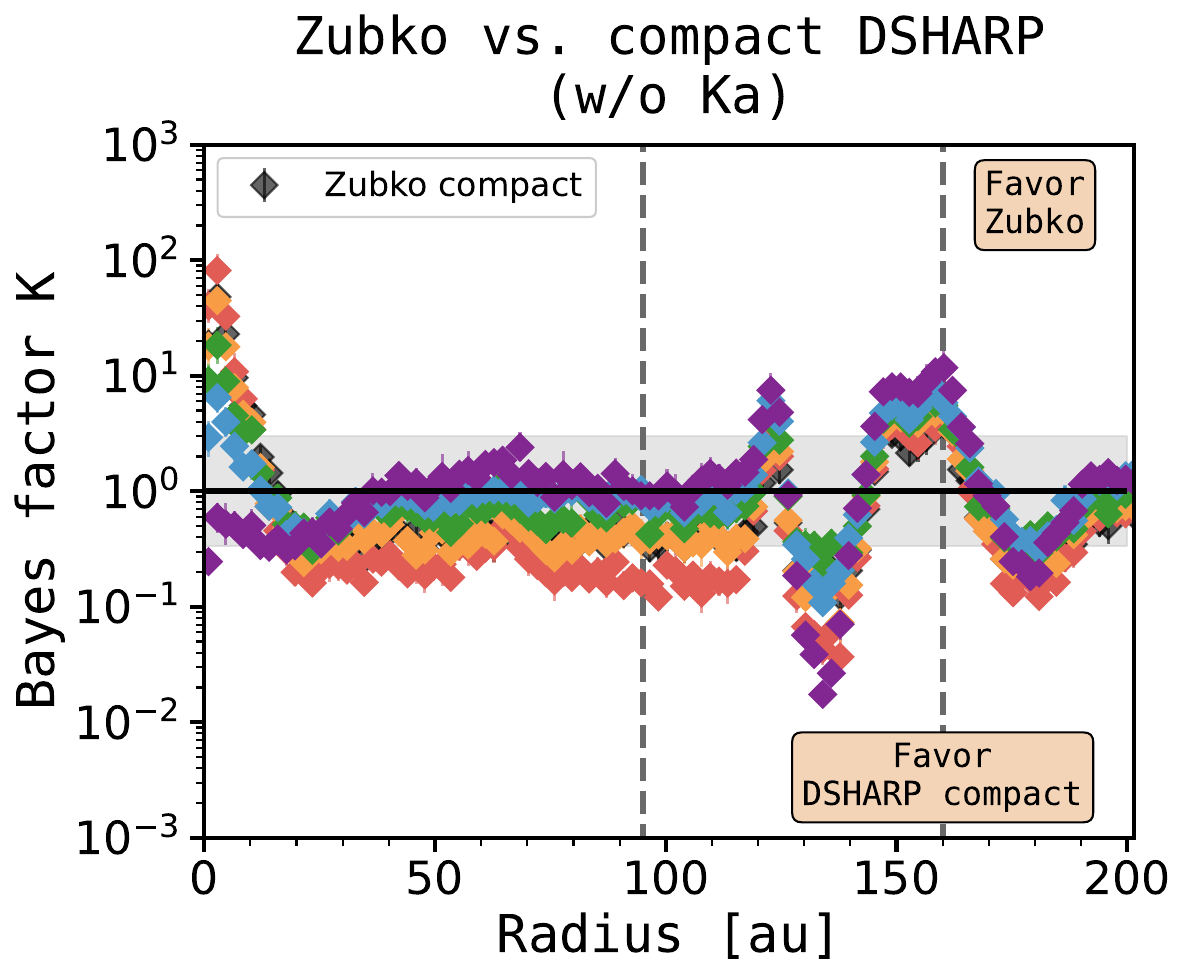}
    \caption{Similar as Figure \ref{fig:dust_compare_bayes}, but for spectrum fitting of $0.18''$-resolution profiles excluding Ka band data. Less pronounced preference on dust mixtures is indicated, highlighting the necessity of long wavelength (broad-band) observation to distinguish dust composition.}
    \label{fig:dust_compare_bayes_noKa}
\end{figure*}

In Section \ref{subsec:model_dust_compare}, we have shown that 90\% porous Zubko mixture has Bayesian evidence similar with compact DSHARP at most radii.
However, in the inner 10 au, there is a preference for less porous Zubko mixtures than compact DSHARP (right panel in Figure \ref{fig:dust_compare_bayes}), although potential spatially resolved non-dust emission might affect the significance of this preference (Section \ref{subsec:result_profiles}).
The HL Tau disk shows a similar preference: when modeling the fraction of amorphous carbon in dust mixtures as a free parameter, \citet{ueda25_hltau} have found a preference for amorphous-carbon-rich dust in inner 40 au, but organics-rich dust outside.
They also link this radial preference to the higher survival temperature of amorphous carbon \citep[$>1000$ K,][]{gail01} than of refractory organics \citep[$300-500$ K,][]{nakano03}, and the preference for amorphous carbon at large radii could be due to a previous accretion outburst.
Larger samples with spatially resolved studies of dust mixture preference are needed to see if this is a common feature among protoplanetary disks.

Across most radii of the MWC 480 disk, the DSHARP mixture is not distinguishable from the Zubko mixture.
The distinct porosities (compact for DSHARP, 90\% for Zubko) might offer a chance to tell them apart with multi-wavelength full polarization observations, since porous grains should produce a flatter spectrum of polarization fraction than non-porous grains \citep[e.g.,][]{tazaki19}.
\citet{harrison24} presented polarization observations of MWC 480 disk at 0.87 and 3 mm and found that polarization patterns at both wavelengths are consistent with dust scattering and a higher polarization fraction at 3 mm in the disk center.
Detailed joint modeling of multi-wavelength continuum and polarization \citep[e.g.,][]{zhang23_hltau} is promising for better constraints on dust composition and porosity, which we refer to future work.

Finally, we underscore the importance of having long wavelength observations.
To evaluate the influence of Ka band, we performed spectrum fitting to radial profiles only from Band 7 to Band 3 at $0.18''$ resolution using the 12 dust mixture models described in Section \ref{subsec:model_dust_method}.
Following Section \ref{subsec:model_dust_compare}, we show the comparison of these mixtures using Bayesian factor in Figure \ref{fig:dust_compare_bayes_noKa}.
When the Ka-band data are excluded, DSHARP mixtures across a wide porosity range can fit the ALMA data equally well, and the preference for the highly porous Zubko mixture becomes less pronounced. 
This result points to the value of including observations at wavelengths beyond 9 mm, which is a promising prospect for the next-generation VLA \citep[ngVLA,][]{ngvla} given its observing efficiency.

\subsection{Constant ring width across wavelengths}
\label{subsec:discuss_ring_width}
In Section \ref{subsec:result_ring_width}, we observed that the radial width of the B95 ring is rather constant (within uncertainty) as a function of wavelength rather than decreasing sharply from 1 mm to 9 mm.
The median values of ring width show a slight decreasing trend with wavelengths: from 6.7 au (1 mm) to 6.1 au (3 mm).
Although the ring width measurement at 9 mm may be affected by image fidelity, we nevertheless discuss the implications of a constant or shallow variation.

The observed super-/sub-Keplerian velocity of CO gas at the inner/outer side of the ring \citep{izquierdo23} strongly indicates the presence of a gas pressure bump at the B95 ring.
For dust particles trapped in a Gaussian pressure bump, the steady-state dust distribution assuming only drift-diffusion equilibrium is approximated as $w_{\rm d}\propto {\rm St}^{-1/2}$ \citep[e.g.,][]{dullemond18}, where $w_{\rm d}$ is the width of dust density profile and ${\rm St}$ is the Stokes number. 
If we simply assume that the observing wavelength linearly correlates with the size of dust particles emitting at the corresponding wavelength, we should expect the ring width at 9 mm to be three times smaller than at 1 mm.
Since we are measuring the intensity ring width instead of the real density ring width, the higher optical depth at 1 mm would widen the intensity ring and make the difference even larger.
For example, \citet{sierra25} found that the widths of three rings in the LkCa 15 disk show a linear relation with frequency from ALMA Band 7 to Band 3.
Therefore, the B95 ring around MWC 480 might be in different physical conditions from LkCa 15 rings.

One possible explanation for the constant ring width is: 
the steady-state dust distribution should not be established only through drift and diffusion of each sized-particle on its own, dust collision (fragmentation and coagulation) is also important.
Recently, \citet{yang25} performed simulations that incorporate drift, diffusion, and dust size evolution in gas pressure bumps.
They found that the radial density distribution of different sized dust in the ring is nearly identical due to dust collisions operating on shorter timescale than drift-diffusion.
Hence, the dominance of dust collisions might explain why the width of B95 ring at 9 mm is not much narrower than at 1 mm \citep[see also][]{pinilla25}. 

The other possibility is that the B95 ring is actually composed of unresolved emission.
For example, the dust particles are accumulated in two unresolved rings with a small separation or many unresolved clumps distributed along a narrow band.
These scenarios could make the observed ring widths to be wider than the beam but stay the same across wavelength.

\subsection{Arc features on the B95 ring}

In Section \ref{subsec:result_frank}, significant arc features in the north-east part of the B95 ring are identified in the residuals after subtraction of axisymmetric FRANK models in Bands 7 and 6, but not detected in Bands 3 and Ka.
Such arcs might indicate dust overdensities in the ring formed by vortex trapping \citep[e.g.,][]{barge95, meheut12, zhu14}, or could be of geometric origin where the asymmetry arises from an optically thick far-side of the disk, which is on our line of sight and directly irradiated by the star \citep{guerra-alvarado24, ribas24}.
The two scenarios expect different variations of the arc-to-ring contrast from short to long wavelengths:
in the vortex trapping scenario, the contrast should increase or similar to that at longer wavelengths due to the possible presence of larger dust grains in the vortex \citep[e.g., ][]{li20}, while the geometric origin instead predicts the opposite since the emission becomes more optically thin.

\begin{figure}[htbp]
    \centering
    \includegraphics[width=1.0\linewidth]{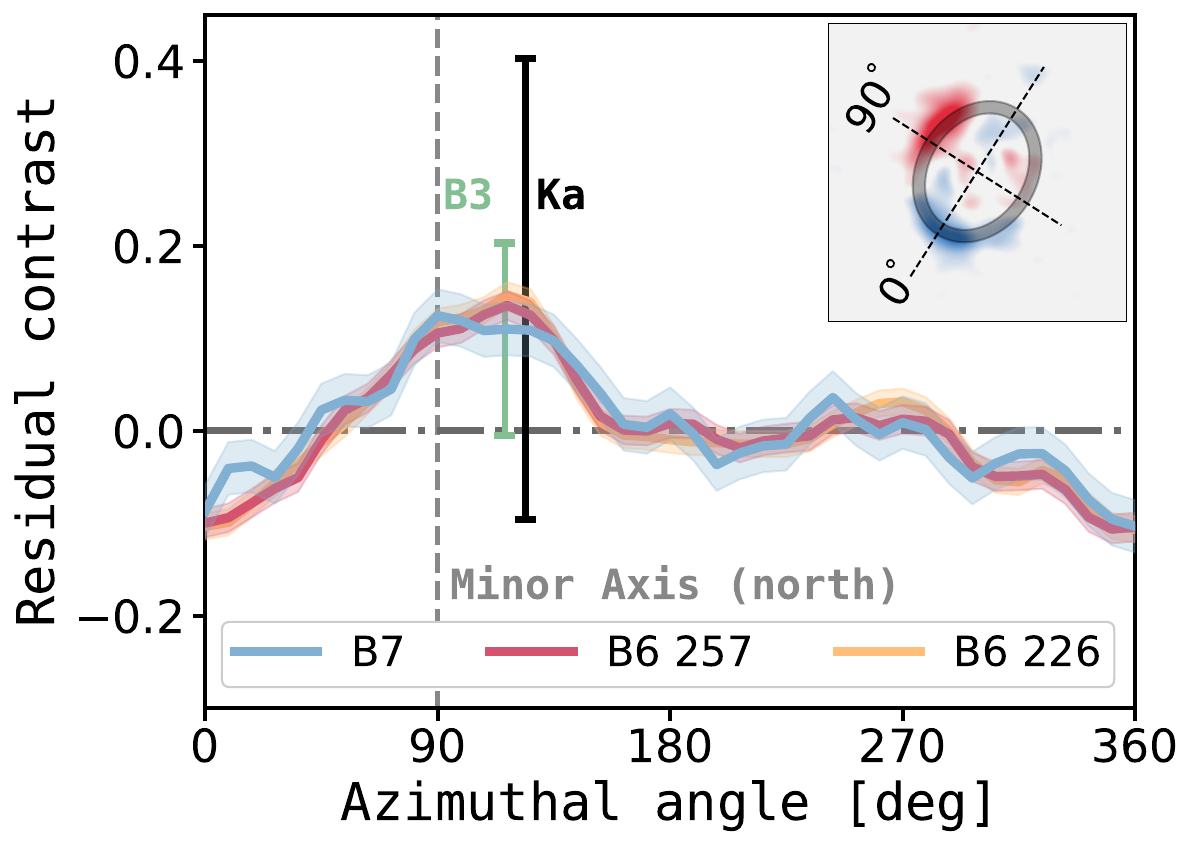}
    \caption{
    Azimuthal profile of residual contrast to ring emission along the B95 ring, obtained from the residual images in Figure \ref{fig:frank_result}. The residual emission is averaged from 0$''$.55 to 0$''$.65 as shown by the black annulus in the upper right inset. For Band 3 and Ka which do not have significant arc-like residuals, only the measured contrast at $120^\circ$ approximately where the arc residuals peak are shown, slight horizontal shifts are applied to separate them.
    }
    \label{fig:arc_profile}
\end{figure}

Figure \ref{fig:arc_profile} presents the azimuthal profile of the residual-to-ring contrast along the B95 ring.
For Band 3 and Ka without significant residuals, the measured contrasts close to the location of the residual peak ($\sim120^\circ$) are shown instead.
The positive arc residuals happen to be on the far-side of the MWC 480 disk \citep[inferred from e.g., $\rm^{12}CO$ channel and moment maps in][]{teague21}, which meets one requirement of the geometric origin.
However, the azimuthal extension is not symmetric about the disk's minor axis, with the peak located on the north side. Multi-wavelength fitting in Section \ref{sec:model_dust} also suggests that the ring emission is optically thin (Figure \ref{fig:dust_opt_depth}).
That said, we cannot rule out the possibility that the B95 ring has an azimuthal variation in dust scale height, which might cause the peak emission to be offset from the disk’s minor axis.
Furthermore, beam smearing in radial profiles could also affect the retrieval of the true dust optical depth of the ring (see the discussion of a possible unresolved substructure in Figure \ref{subsec:result_ring_width}).
Finally, current Band 3 and Ka data lack the sensitivity to measure the arc-to-ring contrast levels and compare them to shorter wavelengths. Further investigation is therefore needed to determine the nature of the arc features.

\subsection{Evidence of planet formation}

\citet{liu19} have demonstrated that a giant planet of mass $\sim2.3~M_{\rm J}$ can carve out the D73/B95 ring/gap substructure around MWC 480 through hydrodynamic simulations.
Most chemical rings are spatially coincident with the gap \citep{law21, jiang22}, which could be produced by a local elevation of C/O ratio by an accreting giant planet \citep{jiang23}.

The shoulders outside B95 ring add more supporting evidence to the presence of a gap-opening planet.
Shoulder features outside dust rings have been reported \citep[e.g.,][]{huangJ20, jennings22_dsharp, jennings22_taurus, yamaguchi24, sierra25_pds70}.
Through 3D multi-fluid hydrodynamic simulations, \citet{bi24} showed that the ring-shoulder pair could be caused by an effect of dust filtration  due to outward gas flow by a massive planet: sub-millimeter grains are entrained by the gas flow out of the pressure maximum, while centimeter grains are not.
It is supported by recent detection of shoulder only at short wavelength in the PDS 70 disk \citep{sierra25_pds70}.
In the case of MWC 480 disk, we have only found shoulder features at Bands 7 and 6 but not at longer wavelengths (Section \ref{subsec:result_frank}), which is consistent with dust filtration induced by planet-driven gas flow.
However, the low-SNR at long baselines (Figure \ref{fig:frank_result}) might prevent the recovery of shoulder feature using FRANK if they are also present at Band 3 and Ka.

Furthermore, the faint ring at $\sim70$ au in the D73 gap, recovered by FRANK at Band 6 (Figure \ref{fig:frank_result}), is reminiscent of the horseshoe dust ring caused by 1:1 gravitational resonant interaction with a planet, which could emerge when a low-mass planet (super-Earth) opens a gap in low-viscosity ($\alpha\leq 10^{-4}$) disks \citep{zhu14, dong17, dong18}.
The growing evidence for a gap-opening planet in the D73 gap underscores the need for deeper, higher-spatial-resolution observations of the MWC 480 disk. 
Definitive detection of both the shoulder and the faint ring within the gap would place stringent constraints on the planet mass and the disk viscosity.

\section{Summary} \label{sec:summary}

Previous millimeter-wavelength imaging of MWC 480 has revealed that its circumstellar disk comprises a smooth, bright inner core, a prominent ring at $\sim95$ au (B95), and a faint outer ring at $\sim160$ au (B160). The B160 ring has been detected in ALMA Bands 7 and 6, but not in Band 3 \citep{long18, law21, sierra21, harrison24}.

In this paper, we present, for the first time, deep VLA Ka Band (9.1 mm) continuum observations of the protoplanetary disk around the Herbig Ae star MWC 480 in Taurus.
By combining these data with archival ALMA observations in Bands 7, 6, and 3, we conduct a detailed investigation of the dust properties in this disk.
We summarize our main results as follows:

\begin{enumerate}
    \item We detected the inner core and the B95 ring in Ka Band. The combination of archival programs revealed the presence of B160 ring at Band 3.
    A point source at the disk center at Ka Band is detected through the constant visibilities at long baselines, which contributes $\sim12\%$ to the total Ka flux.
    This point-like emission could be due to unresolved free-free and gyro-synchrotron emission close to the central star.
    \citet{painter25} estimated instead a $\sim25\%$ fraction of flux at Ka to be non-dust emission, which could be the presence of an extended non-dust emission at Ka such as a jet or ionized photoevaporative wind.
    
    \item Through non-parametric modeling of visibilities using \texttt{FRANK} \citep{frank}, we identified a plateau between 20 and 50 au at all wavelengths with the exact domain depending on the wavelength. 
    At Band 6, the plateau is split into two consecutive smaller plateaus.
    The plateau coincides spatially with a few chemical rings \citep[$\rm CH_3CN$ and $\rm HC_3N$,][]{law21}.
    Outside the B95 ring, \texttt{FRANK} recovered shoulder features at Band 7 and 6. The shoulder at Band 6 is consistent with previous work by \citet{jennings22_taurus} and \citet{yamaguchi24}.
    In the D73 gap, a small faint ring is revealed at Band 6, which is consistent with \citet{jennings22_taurus} which used a different dataset.
    The ring-shoulder pair and the faint ring in the gap add more evidence to the presence of a gap-opening planet.

    \item The B95 ring shows arc-like residuals in the north east in Bands 7 and 6 after subtracting \texttt{FRANK} models. The Band 6 arc was also identified in \citet{andrews24}.
    The arcs happen to be located on the far side of the disk, but their asymmetry about the disk's minor axis and their optically thin emission argue against a purely geometric origin, despite the associated uncertainties.
    Furthermore, the current Band 3 and Ka-band data lack the sensitivity to measure the arc-to-ring contrast and compare it to that at shorter wavelengths. Such a comparison is crucial for distinguishing between a geometric origin and the vortex trapping scenario.

    \item The deconvolved width of the B95 ring is rather constant across wavelengths from 1 mm to 9 mm, which contradicts with that larger dust grains are more radially concentrated in pressure bump.
    We attribute this to evidence that dust collision processes, including fragmentation and coagulation, play a dominant role over drift$-$diffusion \citep{yang25}.
    The other possible explanation is that the B95 ring is to be made up of unresolved emission instead of being one dust ring.

    \item We modeled MWC 480 disk's multi-wavelength radial profiles of brightness using nested sampling.
    We tested 12 dust mixture models, including two compositions (DSHARP and Zubko) and six porosities (compact, 10\%, 30\%, 50\%, 70\% and 90\%).
    Zubko mixture replaces the refractory organics in DSHARP mixture with amorphous carbon, which generally results in higher absorption opacities.
    By comparing Bayesian evidence of each mixture model, we found that both the compact DSHARP and 90\% porous Zubko dust mixture can produce the observed intensities very well at most radii, which are adopted as our fiducial dust mixtures.
    The inner 10 au favors Zubko dust with porosity $\leq70\%$, which is similar to the HL Tau disk \citep{ueda25_hltau}.

    \item The retrieved dust properties show that $a_{\rm max}$ locally peaks around the two rings, and reaches centimeter size in the B95 ring.
    The $a_{\rm max}$ appears constant across the B95 ring, which indicates that it is dominated by turbulent fragmentation \citep{jiang24}, in agreement with the implication from the constant ring width across different wavelengths.

    \item The spectrum modeling shows that inner 10 au is optically thick ($\tau_\nu\geq 1$) at all wavelengths for compact DSHARP mixture, while from 0.9 mm to 1.3 mm for 90\% Zubko mixture.
    The plateau feature from visibility modeling around similar location might be due to the transition from optically thin to thick, so that the emergent intensity becomes insensitive to the increasing dust surface density.

    \item For the two fiducial dust mixtures, the total dust masses (within 200 au) are estimated as
    $860^{+95}_{-78}\rm~M_\oplus$/$1500^{+440}_{-330}\rm~M_\oplus$ for compact DSHARP with large/small-grain solution in inner 30 au, and $230^{+14}_{-13}\rm~M_\oplus$ for 90\% porous Zubko.
    The inner 70 au contains the most dust masses, with fractions of $77\%-93\%$ depending on the dust composition.
    The B95 ring has dust of $100^{+5}_{-5}~M_\oplus$ for compact DSHARP and $43^{+2}_{-2}~M_\oplus$ for 90\% porous Zubko, which implies that there are enough dust materials to assemble the core of giant planets by pebble accretion.

    \item We underscore the value of including long-wavelength observations for multi-wavelength studies on constraining dust mixtures (e.g., composition and porosity).
    In the case of MWC 480 disk, excluding Ka would result in less pronounced preferences for the compact DSHARP mixtures and highly porous ($\sim90\%$) Zubko mixtures.
\end{enumerate}

MWC 480 disk serves as an intriguing case for the study of dust evolution in ring substructures and constraining dust compositions.
Higher spatial resolution and deeper observations (including full polarization) at currently existing and longer wavelengths will be needed to have better constraints on the physical properties, which should be promising prospects for the ngVLA \citep{ngvla}.


\section*{Acknowledgments}

The authors thank the anonymous referee for their constructive comments and suggestions, which helped improve the clarity and quality of this manuscript.
The authors are grateful to Rachel Harrison and Ian Stephens for generously sharing the calibrated and reduced ALMA Band 7 data used in this work. 
Y.S. is grateful for helpful conversations with Elena Viscardi, Linhan Yang, Francesco Zagaria, Satoshi Okuzumi and Til Birnstiel.
Y.S. and G.J.H. are supported by general grants 12573031 and 12173003 from the National Natural Science Foundation of China.
Y.S. acknowledges support from the ESO Studentship Programme.
P.P. acknowledges funding from the UK Research and Innovation (UKRI) under the UK government’s Horizon Europe funding guarantee from ERC (under grant
agreement No 101076489).
D.H. is supported by the Ministry of Education of Taiwan (Center for Informatics and Computation in Astronomy grant and grant number 110J0353I9) and the National Science and Technology Council, Taiwan (Grant NSTC111-2112-M-007-014-MY3, NSTC113-2639-M-A49-002-ASP, and NSTC113-2112-M-007-027).
This research was enabled in part by support provided by the High-performance Computing Platform of Peking University.

This paper makes use of the following ALMA data: ADS/JAO.ALMA\#2016.1.01042.S, ADS/JAO.ALMA\#2017.1.00470.S, \newline ADS/JAO.ALMA\#2018.1.01055.L. 
ALMA is a partnership of ESO (representing its member states), NSF (USA) and NINS (Japan), together with NRC (Canada), NSTC and ASIAA (Taiwan), and KASI (Republic of Korea), in cooperation with the Republic of Chile. The Joint ALMA Observatory is operated by ESO, AUI/NRAO and NAOJ. 
The National Radio Astronomy Observatory is a facility of the National Science Foundation operated under cooperative agreement by Associated Universities, Inc.

%

\vspace{5mm}
\facilities{VLA, ALMA}


\software{\texttt{analysisUtils} \citep{analysisUtils}, 
        \texttt{CASA} \citep{casa2022},
        \texttt{dsharp_opac} \citep{birnstiel18},
        \texttt{frank} \citep{frank},
        \texttt{GoFish} \citep{GoFish},
        \texttt{UltraNest} \citep{ultranest}
          }



\appendix
\restartappendixnumbering
\section{Image improvement by self-calibration}
\label{appsec:self_cal}
\begin{figure*}[htbp]
    \centering
    \includegraphics[width=1.0\linewidth]{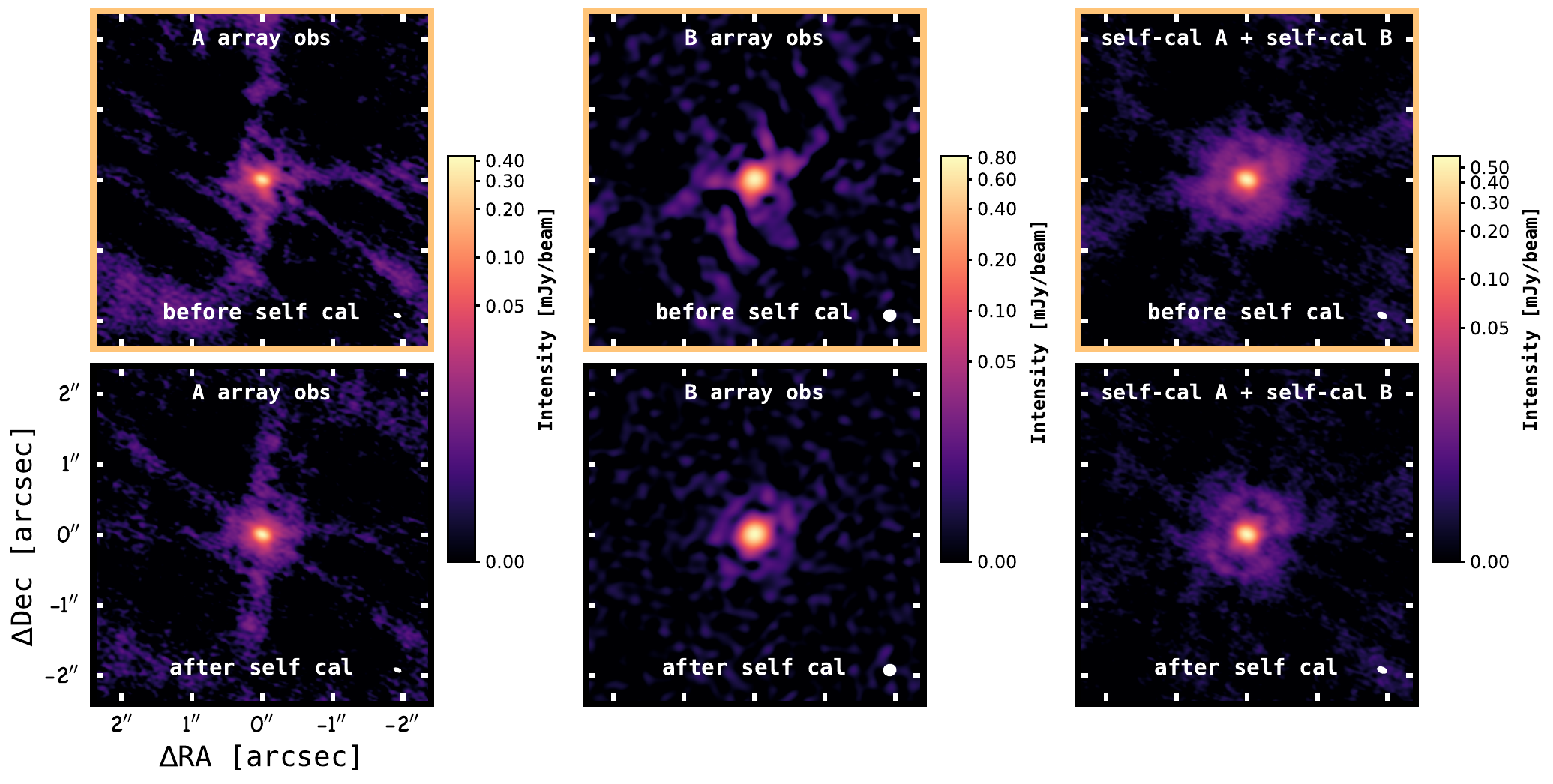}
    \caption{Comparison of Ka-band continuum images before (the upper row) and after (the lower row) phase-only self-calibration. The left and middle columns are for A-array and B-array data, respectively. The right column starts with the data concatenated with self-calibrated A-array and B-array data (the lower left two panels), then another round of self-cal is applied to the concatenated dataset.
    The lower right image is the final product.
    }
    \label{fig:app_selfcal_compare}
\end{figure*}

The self-calibration procedures and peak SNR improvements are described in Section \ref{subsec:obs_vla}, here we detail more on how images are exactly improved after each step. Figure \ref{fig:app_selfcal_compare} shows the continuum images before and after self-calibration. The self-calibration starts with A-array data and B-array data separately. The A-array image shows a broad and faint patch southeast to our target, and the ring is challenging to distinguish from striped artificial emission. After self-calibration, the southeast patch is clearly suppressed and the tenuous ring is more in shape and visible. B-array image experience the largest improvement in the emission morphology: the ring is not visible at all despite that the inner bright core is nicely recovered, instead there are a few `tentacles' connecting to the inner disk. Single round phase-only self-calibration saves the ring from these `tentacles'. Then we combined the self calibrated A-array and B-array dataset, while the image quality is still not ideal: the ring appears more radially extended in the southeast and northwest positions (a bit fluffy-like) and the gap is kind of bridged at these position angles. Hence, we went for another phase-only self-calibration on this concatenated dataset and successfully removed these artificial features and finally had the ring and gap recovered reasonably well. 

\section{Imaging results}
\input{table_imaging_results}
We summarize our continuum imaging results in Table \ref{tab:image_properties}.

\section{Deconvolved width of the B95 ring}
\label{appsec:ring_width}
We summarize our measurement results of the deconvolved width of the B95 ring in Table \ref{tab:ring_fit} as described in Section \ref{subsec:result_ring_width}.

\begin{deluxetable*}{ccccccccccc}
    \label{tab:ring_fit}
    \tablecaption{Gaussian fitting results of the B95 ring}
    \tablehead{
    \colhead{Band} & \colhead{Fit range} & \colhead{$A$} & \colhead{$r_0$} & \colhead{$\sigma$} & \colhead{$\sigma_b$}  & \colhead{$\sigma/\sigma_b$} & \colhead{$w_d$} \\
    \colhead{ } & \colhead{[au]} & \colhead{[$\rm Jy\cdot sr^{-2}$]} & \colhead{[au]} & \colhead{[au]} & \colhead{[au]} & \colhead{} & \colhead{[au]} 
    }
    \tablecolumns{8}
    \startdata
    \cutinhead{Fitting ring profile @ $0.18''$ beam}
    B7 & 85.9 - 109.3 & $1.07^{+0.02}_{-0.02}\times10^{10}$ & $93.0^{+0.8}_{-1.1}$ & $17.3^{+1.6}_{-1.3}$ & 13.3 & $1.30^{+0.12}_{-0.10}$ & $11.0^{+2.4}_{-2.2}$
    \\
    B6 257GHz & 85.9 - 109.3 & $4.85^{+0.07}_{-0.07}\times10^{9}$ & $93.0^{+0.7}_{-0.9}$ & $16.5^{+1.4}_{-1.1}$ & 13.3 & $1.24^{+0.11}_{-0.09}$ & $9.8^{+2.2}_{-2.1}$
    \\
    B6 226GHz & 85.9 - 109.3 & $4.07^{+0.06}_{-0.05}\times10^{9}$ & $93.2^{+0.7}_{-0.8}$ & $16.3^{+1.3}_{-1.1}$ & 13.3 & $1.22^{+0.10}_{-0.08}$ & $9.3^{+2.2}_{-2.0}$
    \\
    B3 & 81.2 - 117.1 & $3.82^{+0.10}_{-0.09}\times10^{8}$ & $93.1^{+0.6}_{-0.7}$ & $15.4^{+0.9}_{-0.8}$ & 13.3 & $1.16^{+0.07}_{-0.06}$ & $7.8^{+1.7}_{-1.7}$
    \\
    Ka & 82.8 - 107.8 & $1.30^{+0.04}_{-0.04}\times10^{7}$ & $91.3^{+0.8}_{-0.5}$ & $17.3^{+1.9}_{-1.6}$ & 13.3 & $1.30^{+0.14}_{-0.12}$ & $11.0^{+2.8}_{-2.7}$
    \\
    \cutinhead{Fitting ring profile @ $0.12''$ beam} 
    B6 257GHz & 85.9 - 109.3 & $6.44^{+0.08}_{-0.08}\times10^{9}$ & $94.3^{+0.3}_{-0.3}$ & $11.1^{+0.4}_{-0.4}$ & 8.9 & $1.25^{+0.05}_{-0.04}$ & $6.7^{+0.6}_{-0.7}$
    \\
    B6 226GHz & 85.9 - 109.3 & $5.43^{+0.06}_{-0.06}\times10^{9}$ & $94.3^{+0.3}_{-0.3}$ & $11.0^{+0.4}_{-0.4}$ & 8.9 & $1.24^{+0.04}_{-0.04}$ & $6.4^{+0.6}_{-0.6}$
    \\
    B3 & 81.2 - 109.3 & $5.12^{+0.27}_{-0.27}\times10^{8}$ & $94.4^{+0.6}_{-0.6}$ & $10.8^{+1.1}_{-0.8}$ & 8.9 & $1.21^{+0.12}_{-0.10}$ & $6.1^{+1.7}_{-1.6}$
    \\
    Ka & 89.0 - 107.8 & $1.76^{+0.07}_{-0.07}\times10^{7}$ & $92.2^{+1.2}_{-1.1}$ & $12.2^{+1.2}_{-1.1}$ & 8.9 & $1.37^{+0.13}_{-0.13}$ & $8.4^{+1.6}_{-1.8}$
    \\
    \cutinhead{Fitting ring profile @ $0.10''$ beam} 
    B6 257GHz & 85.9 - 109.3 & $7.30^{+0.08}_{-0.08}\times10^{9}$ & $94.4^{+0.2}_{-0.2}$ & $9.6^{+0.3}_{-0.3}$ & 7.4 & $1.30^{+0.04}_{-0.03}$ & $6.1^{+0.4}_{-0.4}$
    \\
    B6 226GHz & 85.9 - 109.3 & $6.16^{+0.07}_{-0.07}\times10^{9}$ & $94.4^{+0.2}_{-0.2}$ & $9.4^{+0.3}_{-0.3}$ & 7.4 & $1.28^{+0.04}_{-0.03}$ & $5.9^{+0.4}_{-0.4}$
    \\
    Ka & 89.0 - 107.8 & $1.92^{+0.08}_{-0.08}\times10^{7}$ & $92.4^{+1.1}_{-1.1}$ & $10.9^{+1.1}_{-1.1}$ & 7.4 & $1.47^{+0.15}_{-0.14}$ & $8.0^{+1.4}_{-1.5}$
    \\
    \cutinhead{Ring widths from FRANK models (no further deconvolution)}
    B7 & 91.0 - 101.2 & $2.62-2.51\times10^{10}$ & 94.8 - 94.5 & 5.3 - 5.8 & $\cdot\cdot\cdot$ & $\cdot\cdot\cdot$ & 5.3 - 5.8
    \\
    B6 257GHz & 88.0 - 99.5 & $1.58-1.55\times10^{10}$ & 94.4 - 94.4 & 3.7 - 3.8 & $\cdot\cdot\cdot$ & $\cdot\cdot\cdot$ & 3.7 - 3.8
    \\
    B6 226GHz & 88.0 - 99.5 & $1.34-1.31\times10^{10}$ & 94.3 - 94.4 & 3.7 - 3.8 & $\cdot\cdot\cdot$ & $\cdot\cdot\cdot$ & 3.7 - 3.8
    \\
    B3 & 91.2 - 100.5 & $1.17-1.03\times10^{9}$ & 94.5 - 95.1 & 4.4 - 5.0 & $\cdot\cdot\cdot$ & $\cdot\cdot\cdot$ & 4.4 - 5.0
    \\
    Ka & 86.0 - 100.5 & $3.00-2.68\times10^{7}$ & 92.3 - 92.9 & 6.1 - 7.0 & $\cdot\cdot\cdot$ & $\cdot\cdot\cdot$ & 6.1 - 7.0
    \\
    \enddata
    \tablecomments{Results from FRANK models using both aggressive (left, $\alpha, w_{\rm smooth}=1.1,10^{-3}$) and conservative (right, $\alpha, w_{\rm smooth}=1.3,10^{-1}$) hyperparameters are shown.}
\end{deluxetable*}

\section{Opacities of different dust mixtures}
\label{appsec:opacity}
\begin{figure*}[htbp]
    \centering
    \includegraphics[width=1.0\linewidth]{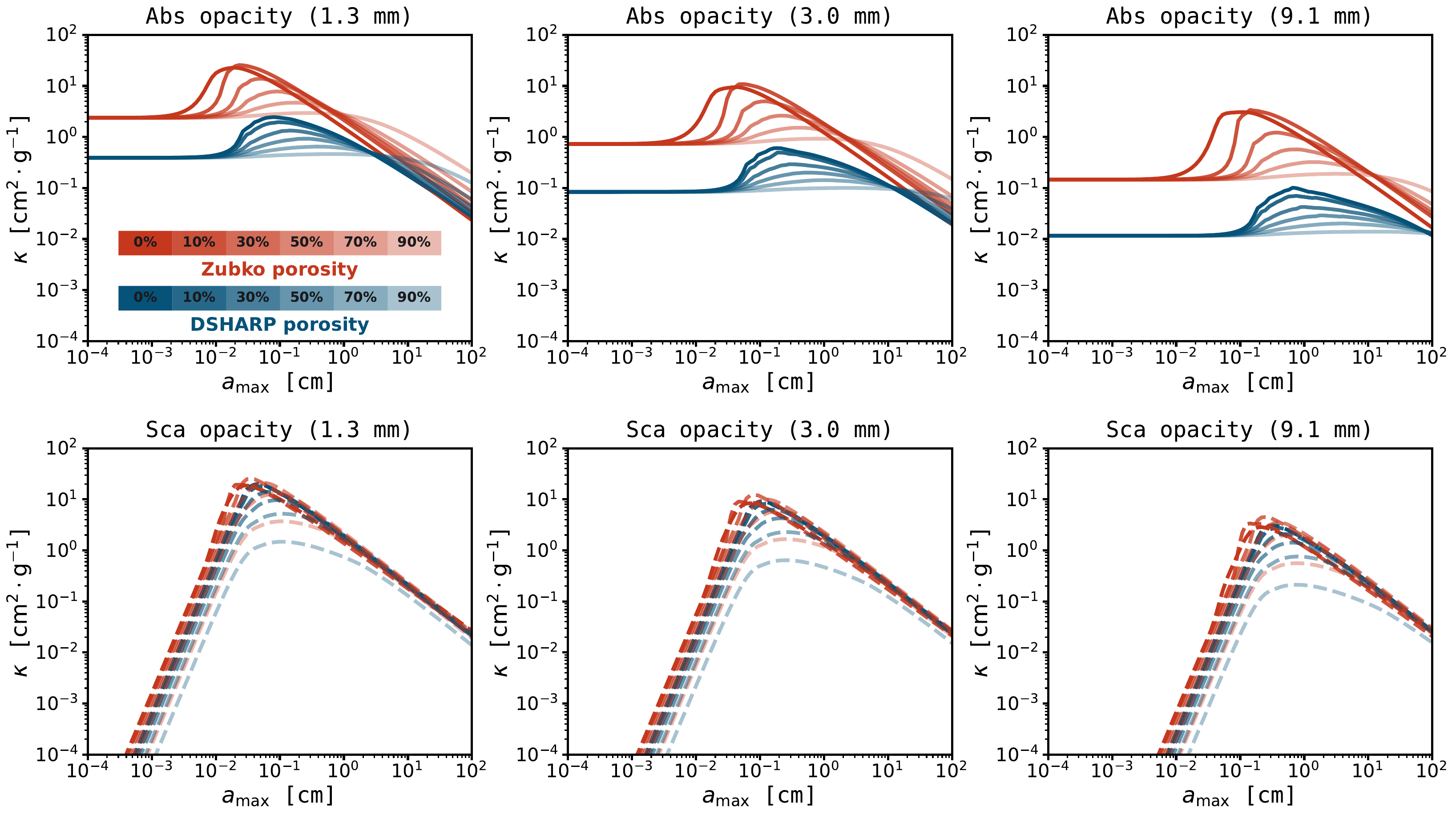}
    \caption{Absorption (upper row) and scattering (lower row) opacities as a function of maximum grain size for DSHARP (blue) and Zubko (red) mixtures with different porosities (with different transparency) at 1.3, 3.0 and 9.1 mm. Dust opacities are derived using a power-law size distribution characterized by an exponent of 3.0.
    }
    \label{fig:app_opacity}
\end{figure*}

Figure \ref{fig:app_opacity} shows the dust absorption and scattering opacities of all dust mixtures in this work at 1.3, 3.0 and 9.1 mm. As described in Section \ref{subsec:model_dust_method}, dust opacities are derived using a power-law size distribution characterized by an exponent of 3.0.

\section{Observed and modeled brightness}
\label{appsec:obs_model_profile}

\begin{figure*}[htbp]
    \centering
    \includegraphics[width=1.0\linewidth]{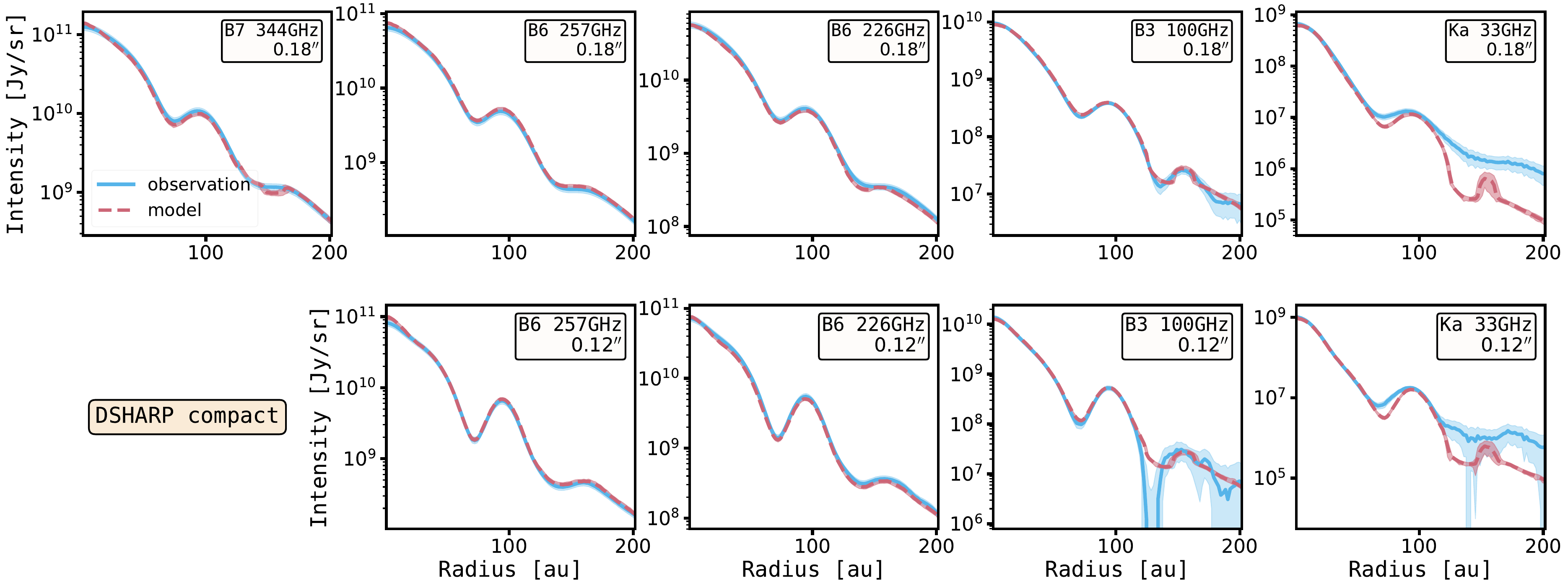}
    \caption{Comparison of intensity profiles between observation and compact DSHARP dust mixture, showing the fitting to both $0.18''$ (upper row) and $0.12''$ (lower row) resolutions.
    The blue solid lines represent observations and the red dashed lines denote the emission from dust mixtures, the corresponding shaded regions display $1\sigma$ uncertainty (including calibration uncertainty).
    }
    \label{fig:app_obs_model_profile_dsharp}
\end{figure*}

\begin{figure*}[htbp]
    \centering
    \includegraphics[width=1.0\linewidth]{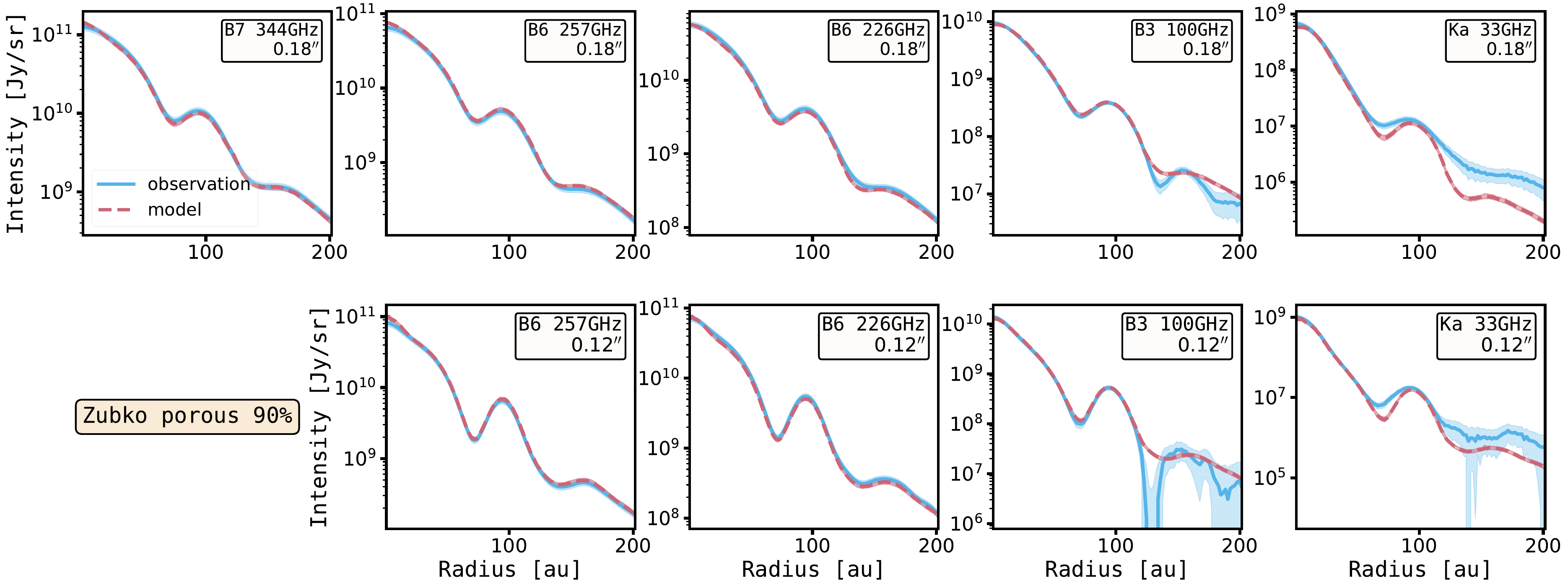}
    \caption{Same as Figure \ref{fig:app_obs_model_profile_dsharp}, but for 90\% porous Zubko dust mixture.
    }
    \label{fig:app_obs_model_profile_zubko}
\end{figure*}

Figure \ref{fig:app_obs_model_profile_dsharp} and Figure \ref{fig:app_obs_model_profile_zubko} compare the intensity profiles between observations and models from different dust mixtures.
Overall, the model emission reproduce the observed intensity quite well within $1\sigma$ uncertainty (including calibration uncertainty).
However, both dust mixtures underpredict the Ka emission in regions of $\sim$ 60 - 95 au and beyond 120 au.
The observed Ka flux in 60 - 95 au is $0.199\pm0.021$ mJy, compared with $0.144\pm0.004$ mJy from compact DSHARP dust and $0.136\pm0.004$ mJy from 90\% porous Zubko dust.
For reference of radius outside 120 au, the observed flux within 120 - 180 au (roughly the outer edge of B160 ring) is $0.068\pm0.010$ mJy, while the modeled fluxes are $0.019\pm0.002$ mJy (compact DSHARP) and $0.029\pm0.001$ mJy (90\% porous Zubko).

\section{Spectrum fitting using MAPS approach}
\label{appsec:reproduce_maps}

\begin{figure*}[htbp]
    \centering
    \includegraphics[width=1\linewidth]{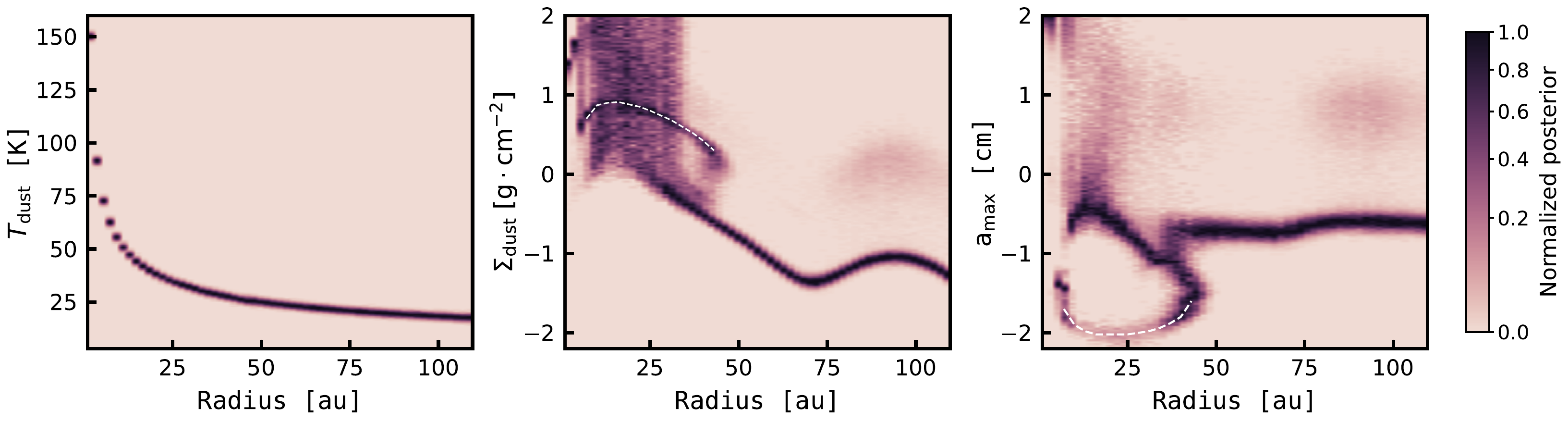}
    \caption{Normalized posterior probability colormap of dust properties, using DSHARP model and mimicking fitting approach in \citet{sierra21}.
    White dashed lines show the branch solution with lower maximum grain size.
    }
    \label{fig:app_dsharp_dust_inner_disk_maps}
\end{figure*}

Figure \ref{fig:app_dsharp_dust_inner_disk_maps} shows the nearly reproduced profiles of dust properties following \citet{sierra21}.
Firstly, the data used for fitting only include ALMA Band 6 (257 and 226 GHz) and Band 3 (here we used data from \citet{sierra21} concatenated with higher-resolution archival data).
Their fitting approach is to use a size distribution $q=2.5$ and fix the dust temperature to the values derived from thermochemical modeling \citep{zhang21_maps}.
We adopt the same $q$ and mimic the fixed temperature by allowing the irradiation temperature to vary within a Gaussian with a width of 1 K.
Our two-branch solution results in Figure \ref{fig:app_dsharp_dust_inner_disk_maps} match quite well with those in \citet{sierra21} (see their Figures 12 and 14).
We retrieve a smaller radial range ($\leq50$ au) where two-branch solutions exist compared to their $\leq$ 60 au, which is likely due to the updated GAIA DR3 distance and our higher resolution profiles used.

\bibliography{ref}{}
\bibliographystyle{aasjournal}

\end{CJK*}
\end{document}

%% file: table_dust_mixture.tex
\begin{deluxetable}{ccc}
    \label{tab:dust_mixture}
    \tablecaption{Dust mixtures used in SED modeling.}
    \tablehead{\colhead{Component} & \colhead{DSHARP (default)} & \colhead{Zubko (BE)}}
    \startdata
    \multirow{2}{*}{\makecell{Water ice$^{(1)}$\\(0.92 $\rm g\cdot cm^{-3}$)}}    & \multirow{2}{*}{0.3642}            & \multirow{2}{*}{0.3642} \\
    & & \\ \hline
    \multirow{2}{*}{\makecell{Astronomical silicates$^{(2)}$\\(3.30 $\rm g\cdot cm^{-3}$)}}     & \multirow{2}{*}{0.1670}            & \multirow{2}{*}{0.1670} \\
    & & \\ \hline
    \multirow{2}{*}{\makecell{Troilite$^{(3)}$\\(4.83 $\rm g\cdot cm^{-3}$)}}     & \multirow{2}{*}{0.0258}            & \multirow{2}{*}{0.0258} \\
    & & \\ \hline
    \multirow{2}{*}{\makecell{Refractory organics$^{(4)}$\\(1.50 $\rm g\cdot cm^{-3}$)}}     & \multirow{2}{*}{0.4430}            & \multirow{2}{*}{$\cdot\cdot\cdot$} \\
    & & \\ \hline
    \multirow{2}{*}{\makecell{Amorphous carbon$^{(5)}$\\(1.80 $\rm g\cdot cm^{-3}$)}}     & \multirow{2}{*}{$\cdot\cdot\cdot$}            & \multirow{2}{*}{0.4430} \\
    & & \\
    \enddata
    \tablecomments{The bulk densities $\rho_{\rm s}$ are 1.675 $\rm g\cdot cm^{-3}$ (DSHARP) and 1.808 $\rm g\cdot cm^{-3}$ (Zubko). The bulk density 1.80 $\rm g\cdot cm^{-3}$ for amorphous carbon follows \citet{optool}. BE means amorphous carbon produced by burning benzene in \citet{zubko96}.}
    \tablerefs{(1) \citet{warren_brandt08}; (2) \citet{draine03}; (3) \citet{henning_stognienko96}; (4) \citet{henning_stognienko96}; (5) \citet{zubko96}.}
\end{deluxetable}

%% file: table_imaging_results.tex
\begin{deluxetable*}{cccllcccccc}
    \label{tab:image_properties}
    \tablecaption{MWC 480 dust continuum imaging results}
    \tablehead{
    \colhead{Band} & \colhead{$\nu$ / $\lambda$} & \colhead{robust} & \colhead{uvtaper} & \colhead{Beam} & \colhead{rms Noise} & \colhead{peak SNR} & \colhead{Flux} & \colhead{$\rho_{\rm eff,68\%}$} & \colhead{$\rho_{\rm eff,90\%}$} \\
    \colhead{} & \colhead{[GHz / mm]} & \colhead{} & \colhead{[mas$\times$mas, deg]} & \colhead{[mas$\times$mas, deg]} & \colhead{[$\rm mJy\cdot beam^{-1}$]} & \colhead{} & \colhead{[mJy]} & \colhead{[au]} & \colhead{[au]}
    }
    \startdata
    \multirow{2}{*}{ALMA B7}  & \multirow{2}{*}{343.5 / \phn0.87}   &  \multirow{2}{*}{-2.0} & 140$\times$20, 97.0 & 179$\times$168, -35.2 & $8.9\times10^{-2}$ & 1152.9 & $776.00\pm1.49$ & $\cdot\cdot\cdot$ & $\cdot\cdot\cdot$ 
    \\
      &  &  & $\hookrightarrow+$\texttt{imsmooth} & 180$\times$180, 0.0 & $8.7\times10^{-2}$ & 1237.6 & $776.01\pm1.50$ & $75.17^{+0.25}_{-0.24}$ &  $119.60^{+0.21}_{-0.20}$
    \\[0.08cm] \hline  
    \multirow{12}{*}{ALMA B6} & \multirow{6}{*}{257.0 / \phn1.17} & \multirow{2}{*}{-1.0} & \multirow{2}{*}{80$\times$60, 95.0} & 96$\times$92, -4.4 & $4.2\times10^{-2}$ & 524.3 & $353.08\pm1.20$ & $\cdot\cdot\cdot$ & $\cdot\cdot\cdot$
    \\
      &  &  &  & 100$\times$100, 0.0 & $4.1\times10^{-2}$ & 591.8 & $353.09\pm1.09$ & $59.71^{+0.18}_{-0.20}$ & $103.72^{+0.08}_{-0.08}$ 
    \\
     &  & \multirow{2}{*}{0.0} & \multirow{2}{*}{86$\times$20, 90.0} & 114$\times$112, 47.9 & $2.1\times10^{-2}$ & 1351.8 & $353.56\pm0.51$ & $\cdot\cdot\cdot$ & $\cdot\cdot\cdot$ 
    \\
      &  &  &  & 120$\times$120, 0.0 & $2.1\times10^{-2}$ & 1497.2 & $353.56\pm0.47$ & $61.23^{+0.19}_{-0.21}$ & $105.10^{+0.06}_{-0.06}$ 
    \\
     &  &  \multirow{2}{*}{0.5} &  \multirow{2}{*}{135$\times$72, 88.0} & 176$\times$168, -87.0 & $2.0\times10^{-2}$ & 2617.8 & $353.75\pm0.31$ & $\cdot\cdot\cdot$ & $\cdot\cdot\cdot$ 
    \\
      &  &  &  & 180$\times$180, 0.0 & $2.1\times10^{-2}$ & 2739.9 & $353.74\pm0.30$ & $68.71^{+0.26}_{-0.27}$ & $113.26^{+0.14}_{-0.14}$
    \\[0.05cm] \cline{2-10}
    & \multirow{6}{*}{226.1 / \phn1.33} & \multirow{2}{*}{-2.0} & \multirow{2}{*}{80$\times$40, 98.0} & 97$\times$93, 44.1 & $5.7\times10^{-2}$ & 345.2 & $294.79\pm1.60$ & $\cdot\cdot\cdot$ & $\cdot\cdot\cdot$ 
    \\
      &  &  &  & 100$\times$100, 0.0 & $5.3\times10^{-2}$ & 398.5 & $294.80\pm1.43$ & $57.93^{+0.21}_{-0.19}$ & $102.90^{+0.10}_{-0.10}$
    \\
     & & \multirow{2}{*}{-0.3} & \multirow{2}{*}{84$\times$20, 98.0} & 111$\times$108, 43.4 & $2.0\times10^{-2}$ & 1208.9 & $295.88\pm0.49$ & $\cdot\cdot\cdot$ & $\cdot\cdot\cdot$ 
    \\
      &  &  &  & 120$\times$120, 0.0 & $1.9\times10^{-2}$ & 1460.2 & $295.88\pm0.43$ & $59.61^{+0.16}_{-0.18}$ & $104.42^{+0.06}_{-0.06}$
    \\
      &  &  \multirow{2}{*}{0.5} & \multirow{2}{*}{130$\times$60, 96.8} & 179$\times$173, -77.4 & $1.6\times10^{-2}$ & 3023.5 & $296.13\pm0.24$ & $\cdot\cdot\cdot$ & $\cdot\cdot\cdot$
    \\
      &  &  &  & 180$\times$180, 0.0 & $1.6\times10^{-2}$ & 3100.5 & $296.13\pm0.24$ & $67.23^{+0.32}_{-0.27}$ & $112.27^{+0.14}_{-0.14}$
    \\[0.08cm]  \hline
     \multirow{6}{*}{ALMA B3} &  \multirow{2}{*}{101.5 / \phn2.95} & \multirow{2}{*}{-0.6} & \multirow{2}{*}{56$\times$5, 100.0} & 118$\times$88, 5.6 & $6.4\times10^{-2}$ & 74.5 & $34.92\pm0.94$ & $\cdot\cdot\cdot$ & $\cdot\cdot\cdot$ 
    \\
      &   &  &  & 120$\times$120, 0.0 & $6.4\times10^{-2}$ & 91.1 & $34.92\pm0.82$ & $48.85^{+0.76}_{-0.79}$ & $101.59^{+0.94}_{-0.88}$
    \\[0.05cm]  \cline{2-10}
      & \multirow{4}{*}{\phn99.8 / \phn3.00}  &  \multirow{2}{*}{-1.2} & $\cdot\cdot\cdot$ & 142$\times$95, 5.4 & $3.3\times10^{-2}$ & 154.9 & $30.34\pm0.50$ & $\cdot\cdot\cdot$ & $\cdot\cdot\cdot$
    \\
      &   &  & \texttt{imsmooth} & 144$\times$100, 5.4 & $3.1\times10^{-2}$ & 171.8 & $30.34\pm0.46$ &  $45.65^{+0.37}_{-0.36}$ & $98.06^{+0.41}_{-0.41}$
    \\
      &   &  \multirow{2}{*}{-0.8} & \multirow{2}{*}{120$\times$2, 94.0} & 178$\times$162, 2.9 & $2.0\times10^{-2}$ & 391.4 & $30.49\pm0.23$ & $\cdot\cdot\cdot$ & $\cdot\cdot\cdot$
    \\
      &   &  &  & 180$\times$180, 0.0 & $1.9\times10^{-2}$ & 430.2 & $30.48\pm0.21$ & $49.43^{+0.32}_{-0.33}$ & $99.93^{+0.28}_{-0.29}$
    \\[0.08cm] \hline
    \multirow{5}{*}{VLA Ka}  & \multirow{5}{*}{\phn33.0 / \phn9.08}  &  \multirow{3}{*}{0.3} & \multirow{3}{*}{60$\times$10, -20.0} & 97$\times$82, 77.2 & $1.6\times10^{-3}$ & 281.5 & $1.63\pm0.03$ & $\cdot\cdot\cdot$ & $\cdot\cdot\cdot$
    \\
      &   &  &  & 100$\times$100, 0.0 & $1.7\times10^{-3}$ & 303.5 & $1.63\pm0.02$ &  $29.75^{+0.50}_{-0.45}$ & $91.44^{+0.71}_{-0.76}$ 
    \\
     &   &  &  & 120$\times$120, 0.0 & $1.7\times10^{-3}$ & 350.8 & $1.64\pm0.02$ &  $30.95^{+0.43}_{-0.37}$ & $91.46^{+0.65}_{-0.68}$ 
    \\
     &   &  \multirow{2}{*}{2.0} & \multirow{2}{*}{90$\times$10, -23.0} & 175$\times$169, 50.1 & $1.6\times10^{-3}$ & 466.5 & $1.64\pm0.03$ & $\cdot\cdot\cdot$ & $\cdot\cdot\cdot$
    \\
     &   &  &  & 180$\times$180, 0.0 & $1.6\times10^{-3}$ & 468.2 & $1.64\pm0.03$ & $35.73^{+0.42}_{-0.40}$ & $92.30^{+0.80}_{-0.87}$
    \\[0.08cm] \hline
    \multirow{2}{*}{VLA X}   & \multirow{2}{*}{\phn10.0 / 29.98} &  0.5  & $\cdot\cdot\cdot$ & 263$\times$199, -73.0 & $2.4\times10^{-3}$ & 71.9 & $0.27\pm0.02$ & $\cdot\cdot\cdot$ & $\cdot\cdot\cdot$
    \\
     &  &  2.0  & $\cdot\cdot\cdot$ & 407$\times$281, -69.5 & $2.1\times10^{-3}$ & 88.7 & $0.27\pm0.02$ & $\cdot\cdot\cdot$ & $\cdot\cdot\cdot$
    \\[0.08cm] 
    \enddata
\end{deluxetable*}